\documentclass[titlepage]{article}
\usepackage{textcomp}
\def\BibTeX{{\rm B\kern-.05em{\sc i\kern-.025em b}\kern-.08em
    T\kern-.1667em\lower.7ex\hbox{E}\kern-.125emX}}

\usepackage[usenames,dvipsnames,svgnames,x11names]{xcolor}
\newcommand{\ysnote}[1]{} % needs a response

\newcommand{\ysnoted}[1]{} % postpone addressing of comment
\newcommand{\aknoted}[1]{}
\usepackage[normalem]{ulem} % required for strikeout font
\newcommand{\para}[1]{{\noindent \textbf{#1.~}}}

\usepackage[nocompress]{cite}
\usepackage{listings}
\lstdefinelanguage{XML}
{
	basicstyle=\ttfamily\footnotesize,
	morestring=[b]",
	moredelim=[s][\bfseries\color{Maroon}]{<}{\ },
	moredelim=[s][\bfseries\color{Maroon}]{</}{>},
	moredelim=[l][\bfseries\color{Maroon}]{/>},
	moredelim=[l][\bfseries\color{Maroon}]{>},
	morecomment=[s]{<?}{?>},
	morecomment=[s]{<!--}{-->},
	commentstyle=\color{gray},
	stringstyle=\color{blue},
	identifierstyle=\color{red}
}
\usepackage{moreverb}

\usepackage[nounderscore]{syntax}

\usepackage[pdftex]{graphicx}
\graphicspath{{./figures/}}
\DeclareGraphicsExtensions{.pdf}
\usepackage[cmex10]{amsmath}
\usepackage{amssymb}
\usepackage{mathtools}
\usepackage{amsthm}
\usepackage{amsfonts}

\usepackage{subfig} %[caption=false,font=footnotesize,labelfont=sf,textfont=sf]
\usepackage{algorithmicx}
\usepackage{algpseudocode}
\usepackage[ruled]{algorithm}
\definecolor{light-gray}{gray}{0.75}
\algrenewcommand{\algorithmiccomment}[1]{\hskip3em{{\footnotesize \textcolor{light-gray}{$\blacktriangleright$}}} #1}
\usepackage{multirow} % Multi-row tables
\usepackage{rotating} % sideways table
\usepackage{booktabs} % better lines
\usepackage{colortbl} % cell colors
\usepackage{tablefootnote} % add support for footnote in table

\usepackage{array}
\definecolor{light-gray}{gray}{0.9} 
\newcolumntype{L}[1]{>{\raggedright\let\newline\\\arraybackslash\hspace{0pt}}m{#1}}
\newcolumntype{C}[1]{>{\centering\let\newline\\\arraybackslash\hspace{0pt}}m{#1}}
\newcolumntype{R}[1]{>{\raggedleft\let\newline\\\arraybackslash\hspace{0pt}}m{#1}}
\newcolumntype{G}[1]{>{\columncolor{light-gray}\centering\let\newline\\\arraybackslash\hspace{0pt}}m{#1}}

\usepackage[pdftex,colorlinks=true,urlcolor=blue,citecolor=blue]{hyperref}

\usepackage{xspace}

\usepackage{enumitem}

\usepackage{blindtext}

\newcommand{\anv}{Anveshak\xspace}
\newcommand{\tf}{TensorFlow\xspace}

\date{}

\begin{document}
	\title{A Scalable Platform for Distributed Object Tracking across a Many-camera Network}
	\author{Aakash Khochare*, Aravindhan Krishnan and Yogesh Simmhan* \\
	~\\
        \emph{*Department of Computational and Data Sciences}\\
        \emph{Indian Institute of Science, Bangalore, India}\\
        ~\\
	\emph{EMail: aakhochare@IISc.ac.in, simmhan@IISc.ac.in}}

	\maketitle
	\begin{abstract}
	Advances in deep neural networks (DNN) and computer vision (CV) algorithms have made it feasible to extract meaningful insights from large-scale deployments of urban cameras. Tracking an object of interest across the camera network in near real-time is a canonical problem. However, current tracking platforms have two key limitations: 1) They are monolithic, proprietary and lack the ability to rapidly incorporate sophisticated tracking models; and 2) They are less responsive to dynamism across wide-area computing resources that include edge, fog and cloud abstractions.  We address these gaps using \emph{Anveshak}, a runtime platform for composing and coordinating distributed tracking applications. It provides a domain-specific dataflow programming model to intuitively compose a tracking application, supporting contemporary CV advances like query fusion and re-identification, and enabling dynamic scoping of the camera network's search space to avoid wasted computation.	We also offer tunable batching and data-dropping strategies for dataflow blocks deployed on distributed resources to respond to network and compute variability. These balance the tracking accuracy, its real-time performance and the active camera-set size. We illustrate the concise expressiveness of the programming model for $4$ tracking applications. Our detailed experiments for a network of $1000$ camera-feeds on modest resources exhibit the tunable scalability, performance and quality trade-offs enabled by our dynamic tracking, batching and dropping strategies.
	\end{abstract}
	
	\section{Introduction} 
	 The push for smarter and safer cities has led to the proliferation of video cameras in public spaces. Regions like London, New York, Singapore and China~\cite{anantha:2017,londoncctv} have deployed camera networks with 1000's of feeds to help with \emph{urban safety}, e.g., to detect abandoned objects~\cite{porikli2008robust}, to track missing people~\cite{arroyo2015expert} and for behavioral analysis~\cite{ko2008survey}. They are also used for \emph{citizen services}, e.g., to identify open parking spots and count the traffic flow~\cite{nafi2012vanet}. Such ``many-camera networks'', when coupled with sophisticated Computer Vision (CV) algorithms and Deep Learning (DL) models~\cite{lecun2015deep}, can also serve as \emph{meta-sensors} to replace other physical sensors for IoT applications and to complement on-board cameras for self-driving cars~\cite{khochare2017distributed}. 
	
\begin{figure}[t]
		\centering
		\includegraphics[width=0.8\columnwidth]{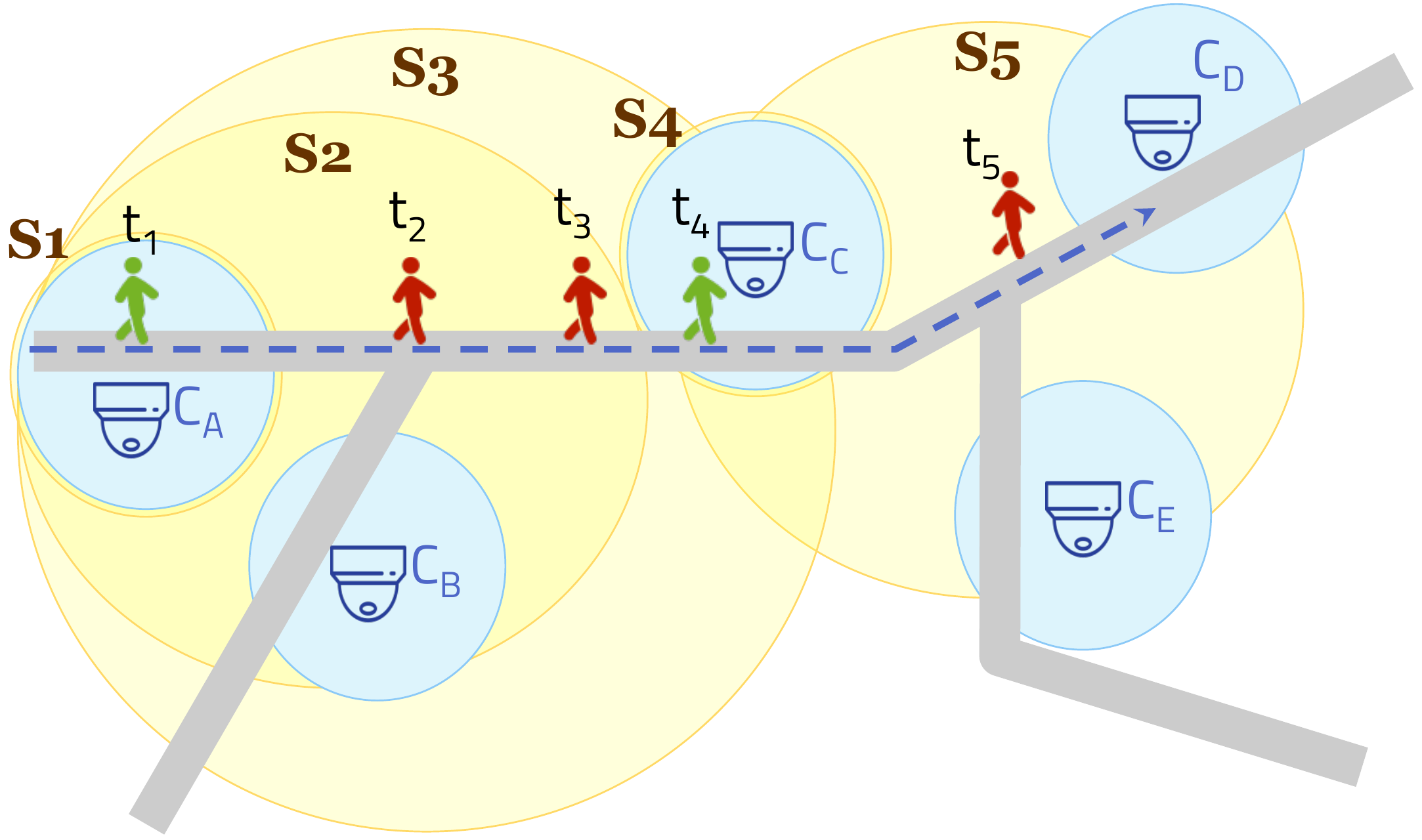}
		\caption{\emph{Spotlight} strategy for camera activation while tracking. Blue circles are the FOV of cameras $C_A$--$C_E$. The person icon shows the location of the person at times $t_1$--$t_5$ on the road; a red icon means they are in a blind-spot while a green icon means they are in the FOV of a camera. The yellow circles $S_i$ are the calculated spotlight regions that indicate the cameras that should be active at time $t_i$.}
		\label{fig:CanonicalExample}
	\end{figure}

	One canonical application domain that operates over such ubiquitous video feeds is called \emph{tracking}~\cite{bedagkar2014survey}. Here, the goal is to \emph{identify an ``object''} or ``entity'' (e.g., a stolen vehicle or a missing child), based on a given sample image or feature vector, in video streams arriving from cameras distributed across the city, and to \emph{track that entity's movements} across the many-camera network in near real-time~\cite{survey-tmcca-2015}. Fig.~\ref{fig:CanonicalExample} illustrates a \emph{missing person} being tracked across a network of 5 video cameras, $C_A$--$C_E$, on a road network using a smart \emph{spotlight} tracking algorithm. A blue circle indicates the \emph{Field of View (FOV)} of a camera. The path taken by the person between time $t_1$ and $t_5$ is indicated by the blue dashed arrow. Given an image of the person, the goal is to locate and trace the path taken by them across the region with high accuracy, while reducing the application design and computing overheads. These poses several challenges.
	
	\textbf{Challenge 1 (Composability).~}
	The application requires online video analytics across space and time, and this commonly has three stages: \emph{object detection, object tracking,} and \emph{re-identification}~\cite{bedagkar2014survey}. The first filters out objects that do not belong to the same class as the entity while the second follows the motion of objects in a single camera's frame~\cite{szegedy2017inception }. 
	Re-identification (or re-id) matches the objects in a camera with the given target entity~\cite{liu2016large}.
	Recently, a fourth stage called \emph{fusion} enhances the original entity query with features from the matched images that is then used for tracking, giving better accuracy~\cite{murthy2018deep}. 
	
	Each of these individual problems is well-researched. But these stages have to be \emph{composed} as part of an overall platform, and coupled with a \emph{distributed tracking logic} that operates across the camera network and over time.
	However, contemporary many-camera analysis platforms are \emph{monolithic, proprietary, and bespoke}~\cite{siebel2004advisor,lim2014isurveillance}.
	They offer limited composability and reusability of models, and minimal support for custom tracking strategies. This increases the \emph{time and effort} to incorporate domain intelligence and adopt the rapid advances being made in CV/DL.
	
	\textbf{Challenge 2 (Distributed Tracking).~}
	It is impractical to execute the full video analytics pipeline on all the cameras due to the punitive computing and network costs. E.g., just doing object detection on a $1000$-camera network using a contemporary fast neural network requires $5$--$128$ Titan XP GPUs, depending on the video resolution and frame-rate, besides the bandwidth to move the video streams to the compute resource~\cite{redmon2017yolo9000}. Instead, these platforms should incorporate \emph{smart tracking strategies} that limit the video processing to the cameras where the object is likely to be present and adapt to \emph{blind-spots}~\cite{esterle2011socio}.
	They can use domain knowledge like the road and transit network, speed of the object, camera location and field of view to make smart choices on the video streams to be actively processed.
	
	\textbf{\emph{Example.~}}
	In Fig.~\ref{fig:CanonicalExample}, the cameras need not generate and process video feeds unless \emph{activated}. Initially, at time $t_1$, the person is within the FOV of $C_A$, and only this camera is made active. By time $t_2$, they have moved out of the FOV of $C_A$, and also of all other cameras, i.e., in a \emph{blind-spot}. Now, we calculate a \emph{spotlight} neighborhood around the camera in which they were last seen, and \emph{activate} cameras that fall within this region, as shown by the yellow circle $S_2$, which overlaps with $C_A$ and $C_B$. This spotlight grows in size to $S_3$ at time $t_3$ as the person is still in a blind-spot, and it activates camera $C_C$ as well. The person reappears in the FOV of $C_C$ at time $t_4$ and the spotlight shrinks to $S_4$ with just this single camera being active and the rest are \emph{deactivated}. The spotlight again grows at time $t_5$ when the person is lost, and $S_5$ activates cameras $C_C, C_D$ and $C_E$. $ \hfill \blacksquare$

	Using such a smart tracking logic to scope the region of interest can significantly reduce the number of active video streams we process, e.g., to 1--3 cameras rather than all 5, in Fig.~\ref{fig:CanonicalExample}. This reduces the resource usage substantially with limited impact on the tracking accuracy. However, contemporary many-camera analysis platforms do not offer such sophisticated and customizable tracking logic. 
	
	\textbf{Challenge 3 (Edge computing and Scaling).~}
	Smart cities are seeing \emph{edge and fog computing resources} being deployed on Metropolitan Area Networks (MAN), to complement \emph{cloud resources}~\cite{varshney:spe:2019}. This brings processing closer to the data source and conserves network bandwidth~\cite{bonomi:2011,lopez:2015}.
	This is important for video tracking, given its low latency, high bandwidth and high compute needs~\cite{satya:pervasive:2015,anantha:2017}.
	So tracking platforms must effectively use such heterogeneous, wide-area compute resources that are part of the computing continuum rather than rely exclusively on cloud resources. 
	
	For \emph{scalability}, the platform must balance the latency for tracking against the throughput supported for the active camera feeds on the available resources -- a high latency can cause the object to be detected late, and lead to the spotlight region growing larger when the person is missing, while a low throughput can limit the number of cameras that can be active at a time, and increase the chances of losing the person. Also, given the \emph{dynamism} in the network behavior, compute performance and stream rates, the platform must trade-off the accuracy of tracking with the application's performance at runtime. 
	\emph{Current platforms do not offer such tunable adaptivity and scaling}~\cite{survey-tmcca-2015,future-icdsc-2017}.
	
	We make the following specific contributions in this article to address these challenges:
	\begin{enumerate}[itemindent=0cm,labelwidth=0.4cm,labelsep=0cm,align=left]
		\item We propose a novel \emph{domain-specific dataflow model} for current and emerging tracking applications, with functional operators to plug-in different analytics. Uniquely, it has first-class support for \emph{distributed tracking strategies} to dynamically decide the active cameras (\S~\ref{sec:model}). These address Challenges 1 and 2. 
		\item We implement the dataflow model and heuristics in our \emph{\anv platform} to execute across distributed edge, fog and cloud resources (\S~\ref{sec:impl}). Further, it incorporates domain-sensitive heuristics for \emph{dropping and batching frames}, which allow users to tune the accuracy, the latency and the scalability under dynamism (\S~\ref{sec:runtime}). These address Challenge 3.
		\item We illustrate the \emph{flexibility} of the dataflow model using $4$ tracking applications, and offer detailed empirical results across latency, accuracy, camera-set sizes and tracking logic to validate the \emph{scalability and tunability} of our platform (\S~\ref{sec:results}).
	\end{enumerate}
	We complement these with a review of related work in \S~\ref{sec:related} and offer our conclusions in \S~\ref{sec:conclusions}.
	
	\section{A Dataflow Model for Tracking}
\label{sec:model}
We first discuss the features of a generic many-camera infrastructure, and then propose a domain-specific dataflow programming model to compose tracking applications.
\subsection{System Model}
A many-camera infrastructure consists of a set of cameras that are statically placed at specific locations in a city, and each can generate a stream of video observations within its FOV~\cite{survey-tmcca-2015}. The cameras are connected to a MAN, directly or through an edge computing device~\cite{satya:pervasive:2015}. (Accelerated) Fog devices may be co-located with the cameras or within a few network hops of them, while cloud resources are accessible at data centers over the Wide Area Network (WAN)~\cite{varshney:spe:2019}. While the edge and fog are typically captive city resources, cloud resources are available on-demand for a price. These resource classes have heterogeneous capacities, and their performance may \emph{vary over time} due to multi-tenancy.  The bandwidth and latency between devices on the MAN and the WAN can be \emph{dynamic}, depending on the traffic. These can affect the QoS of distributed applications.

Cameras allow remote access to their video streams % at the camera are available for access over this network. At the same time, the cameras also exposea
over the network and expose endpoints to control parameters such as the frame rate, resolution and FOV~\cite{videoedge-sec-2018}. Traditionally, servers in the city's control center or in the cloud would acquire the streams for visualization, real-time analytics and archival. But, moving all video data to the compute incurs high bandwidth, and analyzing all streams in real-time can be compute-intensive. Instead, we propose to move the analytics to the data by using edge and fog devices close to the cameras, complemented by the cloud for control. Hence, the tracking platform must operate on heterogeneous, dynamic and distributed compute and network resources.

\subsection{Domain-specific Programming Model}
\begin{figure}[t]
	\centering
	\includegraphics[width=1.0\columnwidth]{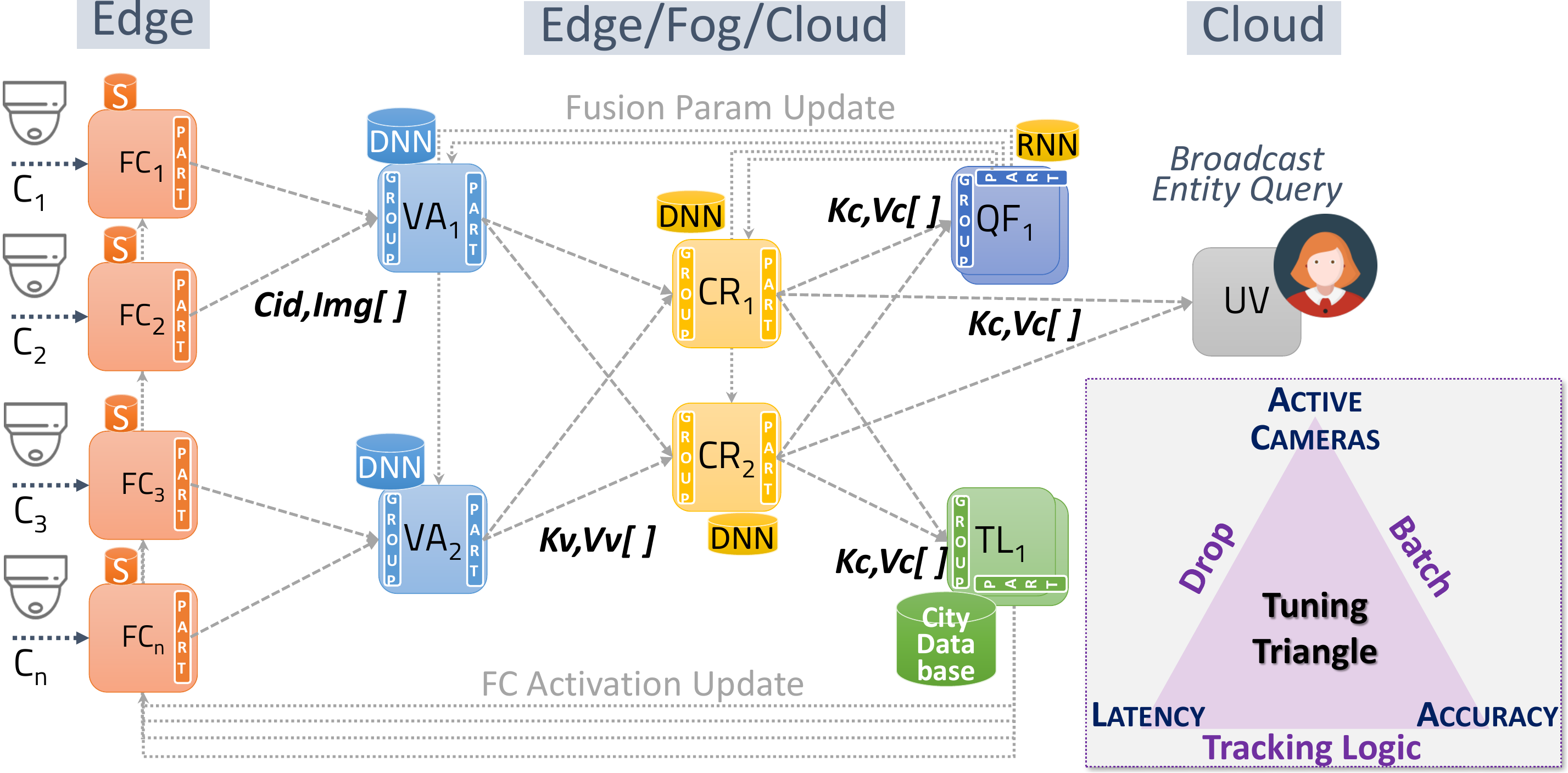}
	\caption{Domain-specific dataflow and modules for tracking. (Inset) Tunable performance and scalability choices.} 
	\label{fig:dataflow}
\end{figure}

We propose a domain-specific model for tracking applications as a \emph{streaming dataflow} with pre-defined \emph{modules} (Fig.~\ref{fig:dataflow}). The user provides the functional \emph{logic} for these modules to compose an application by consuming and producing streams of events (e.g., video frames, detections). We specify the input and output \emph{interfaces} for each module. Multiple \emph{instances} of a module can naturally execute different input events in a data-parallel manner. 

The modules are analogous to the APIs offered by Apache Spark and Hadoop MapReduce~\cite{spark,mapreduce} for pre-defined tasks. However, rather than a general-purpose dataflow like Spark~\cite{spark}, our domain-specific dataflow composition is fixed. This is like MapReduce~\cite{mapreduce} where the user specifies the \emph{Map} and \emph{Reduce} logic, but the dataflow and execution pattern is pre-defined. This eases the development of tracking applications. 
Users can focus on contemporary and emerging advances in DL/CV models that are incorporate into the analytics modules, and uniquely, control the distributed tracking logic through a custom module. Further, our runtime platform offers the benefits of automatic parallelization and performance management.

Next we describe the interfaces of these modules, the dataflow pattern, and the execution model (Fig.~\ref{fig:dataflow}). 

\subsubsection{Filter Controls (FC)} This module is the entry point for video frames from a camera into the dataflow. It is usually co-located with the camera or on an \emph{edge device} connected to it. Each camera has a single FC instance along with its \emph{local state}. Its user-logic decides if a video frame on the input stream of an FC  should be forwarded on its output stream to the Video Analytics (VA) module, or ignored. FC uses its local state (e.g., \emph{isActive}) or even the frame content to decide this. 
If a frame is forwarded, a key-value event is sent on the output stream, with camera ID as key and frame content as value.

Importantly, the FC state for a camera can be \emph{updated} by control events from the Tracking Logic (TL), as described in \S~\ref{sec:dsl:tl}. This allows \emph{tunable activation} of video streams that will enter the dataflow, on a per-camera basis. E.g., TL can have FC deactivate a camera feed if the target will not be present in its FOV, or reduce/increase the frame-rate based on the target's speed. This can balance the dataflow's performance and accuracy. The FC logic should be simple as it runs on edge devices. 

\subsubsection{Video Analytics (VA)} This module receives input event streams from one or more upstream FC modules, and performs \emph{video analytics on a single camera's stream} at a time. This has access to the \emph{entity query} provided by the user, e.g., an image of a person. The user-logic for this module can be complex, and invoke external models in \emph{TensorFlow, PyTorch} or \emph{OpenCV}. The input API for the logic is an iterator of events, \emph{grouped} by the camera ID. This is similar to the \emph{shuffle and reduce} in MapReduce. Grouping by camera ID gives the logic access to a \emph{batch of frames} from the same camera for temporal analytics. This also amortizes the overheads to invoke the external models. We discuss specific batching strategies in \S~\ref{sec:runtime:batch}. Depending on the compute needs of this logic, it may run on edge, fog or cloud resources.

Exemplar VA logic include object detection and tracking using DL models like \emph{YOLOv2}~\cite{redmon2017yolo9000} or classic CV techniques like \emph{Histogram of Gradients (HoG)}. VA can access the user's target query and maintain \emph{local state} across executions, to be passed to the external model. The output of the logic is a batch of key-value pairs, which may be, e.g., the camera ID (key), and bounding boxes for potential target objects in a frame with confidence scores (value). 
There can be a many-to-many relationship between the input and output events for this module. However, we allow users to \emph{link} an output event with an input event, and this provenance lets us trace its latency and help with drop strategies we propose in \S~\ref{sec:runtime:drop} to meet the QoS goals.

Like FC, the local state of this module can be updated by the \emph{Query Fusion (QF)} task. This allows dynamic updates to the entity query by \emph{fusion} algorithms~\cite{murthy2018deep} to enhance a query's feature vector with information from ongoing detections of the entity in the frames. The VA can also update its model based on such signals.

\subsubsection{Contention Resolution (CR)} This module receives a stream of key-value events from one or more VA instances, \emph{grouped by key}. 
The keys are typically the camera ID and the values contain detections or annotated frames, but these can be overridden by the VA user logic. It has access to the entity query as well. This logic is used to analyze results from \emph{multiple cameras}, say, to resolve conflicting detections from different cameras, e.g., the same entity being detected at different parts of the city simultaneously. It can use more advanced re-id logic or DL models for a \emph{higher accuracy match}. CR may be triggered only on a conflict or a low confidence detection by a VA, and hence execute less often than VA, but be compute intensive. CR may even degenerate to a human-in-the-loop deciding on borderline detections. This makes it better suited for running on fog or cloud resources. Like VA, this module can receive updates from QF as well.

The output stream from executing CR primarily contains metadata -- much smaller than the video input -- and this is forked three ways, to TL, QF and UV modules.

\subsubsection{Tracking Logic (TL)} 
\label{sec:dsl:tl}
This is a \emph{novel module} that we propose to help users capture the core logic of distributed tracking across the multi-camera network~\cite{shiva2017distributed}. The detections that TL receives from CR for each frame may be \emph{positive} or \emph{negative}. On a negative detection, the TL logic is expected to \emph{expand} the search space by activating additional cameras, while if the entity is found in a frame (positive), TL should \emph{contract} the search space. The module also use sophisticated tracking algorithms using prior domain knowledge on the environment and the entity, and devise strategies to (de)activate the cameras to optimize the quality and performance of tracking. It can be hosted on cloud resources.

E.g., in Fig.~\ref{fig:CanonicalExample}, TL can use the knowledge of the road network and camera locations to dynamically decide the camera search space (spotlight), depending on when and in which camera the entity was last detected, and (de)activate those cameras. It can also be more sophisticated and change the FOV for cameras to focus on an approaching or receding entity, or change the frame-rate based on the entity's speed. 
This separates the core video analytics logic, from their interpretation over the camera network for distributed entity tracking and camera controls.

\subsubsection{Query Fusion (QF)} This module uses information on the detections to enhance the entity query's features. High-confidence entity detections in the input video can be fused with the existing entity query to generate a new query that offers better matches, or even use negative matches to enhance the query~\cite{shiva2017distributed,murthy2018deep}. The output of this module updates the entity query at the VA and CR modules for their future input streams.  

\subsubsection{User Visualization (UV)} This is a user-facing module that can be used to submit the entity query and display the current state of the tracking and detections. This can be a central portal running on the cloud where authorized personnel can view the progress. 

\subsubsection{Discussion on Event Ordering} The programming model and runtime do not enforce \emph{temporal ordering} of events flowing across different streams in the pipeline. While \emph{temporal ordering} may be required for some vision algorithms such as estimating optical flow~\cite{fleet2006optical}, the user logic can reorder events in a batch using the timestamps passed in the \emph{key}, \emph{value} pairs. The runtime can enforce ordering in the \emph{group} stage of every module using a temporal window and watermarks~\cite{akidau2013millwheel}. However, we currently do not include such mechanisms.

\subsection{Sample Tracking Applications}\label{sec:sampleapp} 
We illustrate several applications designed using our domain-specific dataflow and modules to track an \emph{entity} (e.g., missing person) across a road network. They take an input image as the entity query, and return detections of the entity in the camera network to the UV module in real-time. 
For the first variant (\emph{App 1}), the pseudo-code of the user-logic for different modules are given below. 

\begin{algorithm}[h!]
	\footnotesize
	\caption{Pseudocode of Modules for App 1}
	\label{alg:app1}
	
	\vspace{0.05in}
	\begin{algorithmic}[1]
		\small
		\Procedure{FC}{$img, state$}
		\State \textbf{~~return} $state.get('isActive')$	 
		\EndProcedure	
	\end{algorithmic}
	\hrule\vspace{0.05in}
	
	\begin{algorithmic}[1]
		\small
		\Procedure{VA}{$C_{id}, imgs[~], state$} 
		\State $bbs[~][~]  =$ \Call{OpenCV.HoG}{imgs[~])}
		\For {$img$ in $imgs[~]$ and $outbbs[~]$ in $bbs[~][~]$}
		\State \Call{emit}{$C_{id}, \langle img,  outbbs[~] \rangle)$}
		\EndFor 
		\EndProcedure	
	\end{algorithmic}
	\hrule\vspace{0.05in}
	
	\begin{algorithmic}[1]
		\small
		\Procedure{CR}{$C_{id}, \langle img,  outbbs[~] \rangle[~], state$} 
		\State $query =  state.get('entity\_query\_img')$
		\State $cropped = [~]$
		\For {$tuple$ \textbf{in} $\langle img,  outbbs[~] \rangle[~]$}
		\State $cropped\_img = $ \Call{crop}{$img, outbbs[~]$}
		\State $cropped.append( cropped\_img )$
		\EndFor
		\State $detections =$ \Call{TFlow.DNN\_CR}{$cropped, query$}
		\For {$was\_detected$ \textbf{in} $detections[~]$}
		\State \Call{emit}{$C_{id}, \langle img, was\_detected \rangle$}
		\EndFor
		\EndProcedure	
	\end{algorithmic}
	\hrule\vspace{0.05in}
	
	\begin{algorithmic}[1]
		\small
		\Procedure{TL\_WBFS}{$\langle C_{id}, \langle img, detections[~] \rangle \rangle[~], state$ }
		\State $el = $ \Call{getEntityLocation}{$\langle C_{id}, detections[~] \rangle[~]$}
		\If {$el == \varnothing$} \Comment{Entity lost. Expand spotlight...}
		\State $ graph = state.get('road\_network') $
		\State $lsl = state.get('lastSeenLocation')$
		\State $lst = state.get('lastSeenTime')$
		\State $cameras[~] =$ \Call{weightedBFS}{$graph, lsl, lst$}
		\State \Call{expandSearchSpace}{$cameras$}
		\Else
		\State \Call{shrinkSearchSpace}{$el$}
		\EndIf
		\EndProcedure
	\end{algorithmic}
	\hrule\vspace{0.05in}
	
	\begin{algorithmic}[1]
		\small
		\Procedure{QF}{$\langle C_{id}, \langle img, detections[~] \rangle [~]\rangle, state$} 
		\State $oldFeature \gets state.get('state')$
		\For {$image$ \textbf{in} $img[~]$}
		\If {$detection == true$}
		\State $newFeature \gets RNN(image, oldFeature)$
		\EndIf
		\EndFor
		\State $emit(C_{all}, image[~], out[~])$
		\EndProcedure	
	\end{algorithmic}
	%\hrule
	\vspace{0.05in}
\end{algorithm}

Initially, all FCs have their \emph{active} state set to \emph{true} to allow their input streams to be passed on, and the entity to be found for the first time. The VA gets a batch of images from a camera, and uses a feature-based OpenCV HoG pedestrian detector~\cite{dalal2005histograms} to put bounding boxes (\emph{bbs}) around persons in each image, and sends these to CR. HoG runs on the entire batch of images.  

CR crops and extracts the image regions in the bounding boxes of frames sent from different cameras, and passes these as a batch to a high-quality OpenReid TensorFlow DNN for pedestrian detection~\cite{openreid}. The DNN detects if the query entity is present in the cropped images, and emits the \emph{true} and \emph{false} matches downstream. UV then displays the matching camera frames.

TL also receives these detections, and has access to the road network, road lengths and camera locations. When the entity is not in matched in any camera, it starts a \emph{Weighted Breadth First Search (WBFS)} on the road network from the last known position of the entity, taking into account the road lengths, the entity's default speed and the time elapsed since the last detection. This identifies the spotlight region where the entity should be present within, and TL contacts the FC of cameras in this region to activate them. Else, if the entity is detected in some camera's frame, the spotlight region contracts to that camera.
Lastly, QF uses an RNN~\cite{murthy2018deep} to enhance the entity query using high-quality hits.

Table~\ref{tab:applications} lists the module logic used in this and $3$ other tracking applications we can compose. We use a different DNN~\cite{ahmed2015improved} in CR for \emph{App 2}. The query may also be for a vehicle based on image rather than license plate, in \emph{App 3}, where we also use DNNs for object detection in VA~\cite{redmon2017yolo9000} and CR~\cite{sochor2018boxcars}.
Here, TL can also be more complex, with awareness of the road lengths and the target's speed. In \emph{App 4}, we can use a Na\"{i}ve Bayes model to give the likelihood of paths that will be taken by the entity to decide the cameras to activate. 

\begin{table}[t]
	\footnotesize
	\setlength{\tabcolsep}{4pt}
	\centering
	\caption{Module mappings for illustrative tracking apps}
	\label{tab:applications}
	\begin{tabular}{c|C{0.95cm}|C{1.6cm}|C{1.6cm}|C{1.25cm}|C{0.7cm}}
		\hline
		\bf App\# & \bf FC & \bf VA & \bf CR & \bf TL & \bf QF \\	\hline\hline
		1 & Active? & HoG~\cite{dalal2005histograms} & Person Re-id~\cite{openreid} & WBFS & -- \\\hline
		2 & Active? & HoG~\cite{dalal2005histograms} & Person Re-id~\cite{ahmed2015improved} & BFS & RNN\cite{murthy2018deep}\\\hline
		3 & Frame Rate & YOLO for Cars~\cite{redmon2017yolo9000} & Car Re-id~\cite{sochor2018boxcars} & WBFS w/ speed & -- \\\hline
		4 & Active? & Person Re-id (Small)~\cite{he2016deep} & Person Re-id (Large)~\cite{szegedy2017inception} & Probabi\-listic & -- \\
		\hline
	\end{tabular}
\end{table}

\section{\anv Platform Implementation}
\label{sec:impl}

\begin{figure}[t]
	\centering
	\includegraphics[width=0.9\columnwidth]{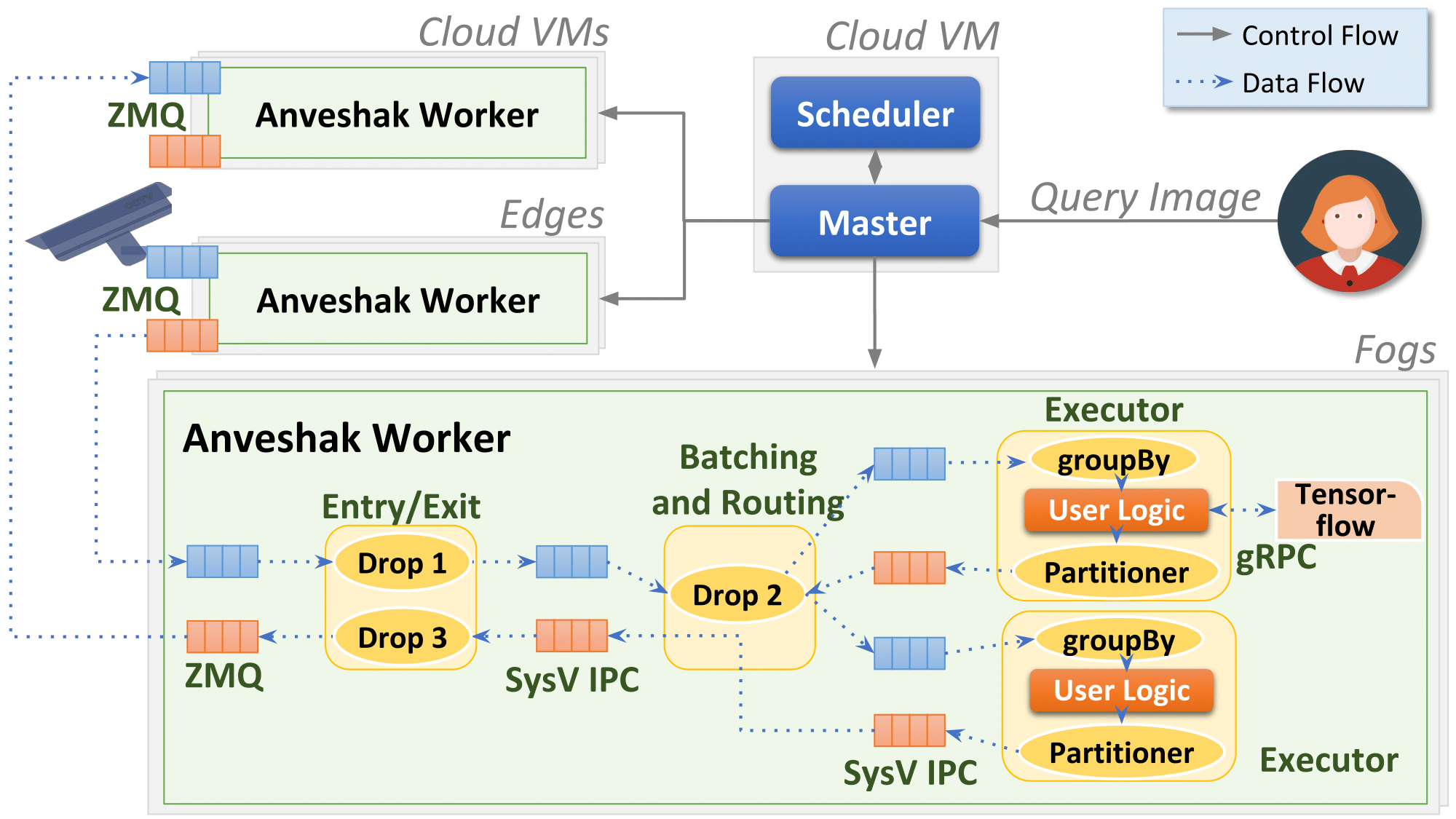}
	\caption{System Architecture of \anv}
	\label{fig:arch}
\end{figure}

We implement this domain-specific dataflow model as \emph{\anv} (\emph{Explorer}, in Sanskrit), a Python-based distributed runtime engine that allows users to easily define their tracking application. Its architecture is illustrated in Fig.~\ref{fig:arch}. \anv is more \emph{light-weight} than Big Data streaming platforms like Apache Spark Streaming or Flink~\cite{spark,carbone2015apache}, and designed to operate on a WAN than a Local Area Network (LAN). This allows it to be deployed on \emph{edge, fog or cloud} resources.

Application developers implement their user logic in Python for the different modules of the dataflow, such as in Table~\ref{tab:applications}. 
External models like TensorFlow are invoked from a module using a local \emph{gRPC} service.
A \emph{Master} process runs in the cloud at a well-known endpoint and manages the application deployment. 
The application definition is submitted to the Master with the module descriptions and their configuration parameters, e.g., the entity query image for VA and CR, or the expected entity speed used by TL. 

The Master calls a \emph{Scheduler} logic that decides the number of module instances and maps them to the resources.
The scheduling logic is modular. By default, we use a simple round-robin scheduler with a fixed number of instances per module type, and map specific module types to specific edge, fog or cloud resource abstractions. More advanced scheduling strategies are beyond the scope of this paper. 

Each distributed resource available for deploying the dataflow runs a \emph{Worker} process, which manages module instances on that resource and transfers data between instances on different devices using \emph{ZeroMQ}~\cite{zmq}. The Master initializes module instances on a resource by contacting its Worker. 
We assume that the required libraries are pre-deployed in the Workers, and in future, this can be replaced by light-weight containers like \emph{minikube}~\cite{minikube} for deployment.

A Worker can host multiple module instances that has the user logic, and each is encapsulated in a separate \emph{Executor} process. We use a \emph{Router} logic to pass data between the Worker and the local Executors using \emph{Sys V} for Inter-Process Communication (IPC)~\cite{bovet2005understanding}. A Worker can fetch the camera feeds from an external endpoint to initiate the dataflow. 

Next, we discuss various dynamic runtime optimization strategies incorporated in the \anv platform to balance the latency performance, detection accuracy and scalability for a tracking dataflow that is executing.

\section{Runtime Tuning Strategies}
\label{sec:runtime}
The \anv platform operates in a dynamic environment, and needs to be tuned at runtime to adapt to these conditions. We offer a novel \emph{Tuning Triangle} (Fig.~\ref{fig:dataflow}, bottom right), where users can achieve two of three \emph{properties}: \emph{performance}, \emph{accuracy} and \emph{scalability} (shown at the corners of the triangle) when performing tracking, and these are controlled by \emph{knobs} (shown at the side opposite to a property's corner). The \emph{batching} knob controls the latency property, the \emph{dropping} knob controls the accuracy, and the sophistication of the \emph{tracking logic} knob, already discussed, determines the active camera set size (or scalability). Next, we discuss the two other knobs to control \emph{data drops} and \emph{batching}.

\subsection{Approach}
We have a captive set of edge, fog and cloud resources having variable compute load due to a changing active set size being processed, and are connected over a MAN/WAN that exhibits dynamism in the latency and bandwidth between resources present on it. So the transient load on resources hosting the active module instances can exceed the available compute or network capacity, which leads to higher event latencies that can cascade up the input event stream. 

In such cases, we can gracefully degrade by \emph{dropping events} that cannot be processed within a \emph{maximum tolerable latency} ($\gamma$) specified by the user.  If we drop potentially stale events early in the dataflow pipeline, we can make make more resources available to the events that are retained and increase their chances of completing within the threshold. This knob helps meet the latency goals and supports a larger active-set size, but it affects the accuracy of tracking if frames containing the entity are dropped. Besides allowing the users to disable dropping, we propose a \emph{smart dropping strategy} in \S~\ref{sec:runtime:drop} to dynamically vary the accuracy, given a tolerable latency and a peak active camera set size.

For timely processing of the video feeds, it is sufficient for the latency between a frame generated at a camera and its processed response reaching the UV to fall within $\gamma$. This can be exploited to enhance the processing throughput by \emph{batching events} passed to the VA/CR modules to amortize the static overheads of invoking the external DL models, while ensuring that the processing latency per event is within permissible limits. However, the time budget available for batching can vary across time, and is non-trivial to estimate without a shared global clock. Besides allowing users to set a fixed batch size, we propose an \emph{adaptive batching strategy} in \S~\ref{sec:runtime:batch} that maximizes the batch size without violating the latency constraint, for a given accuracy requirement and a peak active camera set size.

Data drops and dynamic batching are featured in stream processing systems. Techniques for load shedding (drops)~\cite{abadi2003aurora, abadi2005design} and batching~\cite{lohrmann2015elastic, das2014adaptive} have been proposed to help determine the the fraction of data to be dropped and the batch size. They use greedy empirical approaches or model it as an optimization problem that is solved using numerical solvers. But they make centralized decisions, are computationally costly and/or expect synchronized device clocks. These are challenging on constrained and wide-area distributed resources.
Instead, we design strategies that are lightweight, distributed and resilient to clock-skews. 

\subsection{Preliminaries}
For modeling latency, we decompose the dataflow graph of module instances (tasks) shown in Fig.~\ref{fig:dataflow} to a set of sequential task \emph{pipelines}, with a \emph{task selectivity} of $1$:$1$ -- the ratio of input to output events. Each sequential pipeline comprises of F, VA, CR and VU instances, though we assume these are generic tasks, $[\tau_1, \tau_2, ..., \tau_n]$, where $\tau_1$ is the source task and $\tau_n$ is the sink task. We propose strategies for an individual pipeline, which is then generalized to the entire dataflow.

Each event $e_k$ arriving at the source task $\tau_1$ of each pipeline is assigned a unique ID $k$. This ID propagates to all its causal downstream events. Since we have a $1$:$1$ selectivity, an event $e_k^i$ in the pipeline can be uniquely identified by a combination of its source event ID $k$ and the task $\tau_i$ it is an input to.

When an event $e_k^i$ arrives at a task $\tau_i$ from an upstream task $\tau_{i-1}$, it is placed in a FIFO queue (Fig.~\ref{fig:task}). Events at the front of the queue are identified by the Executor to form a batch, whose size is dynamically decided, as discussed in \S~\ref{sec:runtime:batch}. The user-logic is triggered on the batch of input events and it returns a batch of output events that is passed to a \emph{partitioner}, which routes each event based on its key to a downstream task. Let $a^i_k$ indicate the \emph{arrival time} of an event $e_k^i$ at a task $\tau_i$ from its upstream task (Fig.~\ref{fig:task}). This timestamp is measured at the resource hosting the task $\tau_i$. The time spent by the event in the queue before execution is given by the \emph{queuing duration} $q^i_k$.
Once events from the queue are formed into a batch of size, say $b$, let the function $\xi_i(b)$ give the \emph{estimated execution duration} for the batch by the user-logic for the task $\tau_i$. We assume that the \emph{execution duration} monotonically increases with the batch size, i.e., $\xi(b) < \xi(b+1)$. When $b=1$, this is a streaming execution with no batching delay. We also define the \emph{processing duration} $\pi^i_k = q^i_k + \xi(b)$, as the time between an event arriving at a task and the resulting output event being placed on its output stream. 

We define the \emph{upstream time} for an event $e_k^i$ arriving at task $\tau_i$ as $u^i_k  = a^i_k - a^1_k$. This is a relative time defined using the timestamps of the source event $e^1_k$ at the source task and the causal event $e^i_k$ observed at the current task, which in turn depend on their local device clocks $\kappa_1$ and $\kappa_i$. The arrival time $a^1_k$ for the source event $e^1_k$ is propagated to all its causal downstream events in their headers.

While we initially assume all device clocks are synchronized, we later discuss how our techniques are resilient to clock-skews between all devices (as is common in MAN/WAN), except those hosting the source and sink tasks of the pipeline, $\kappa_1$ and $\kappa_n$~\cite{buchholz2007brief,geng2018exploiting}.

\subsection{Strategies to Drop Events}
\label{sec:runtime:drop}

\begin{figure}[t]
	\centering
	\includegraphics[width=1.0\columnwidth]{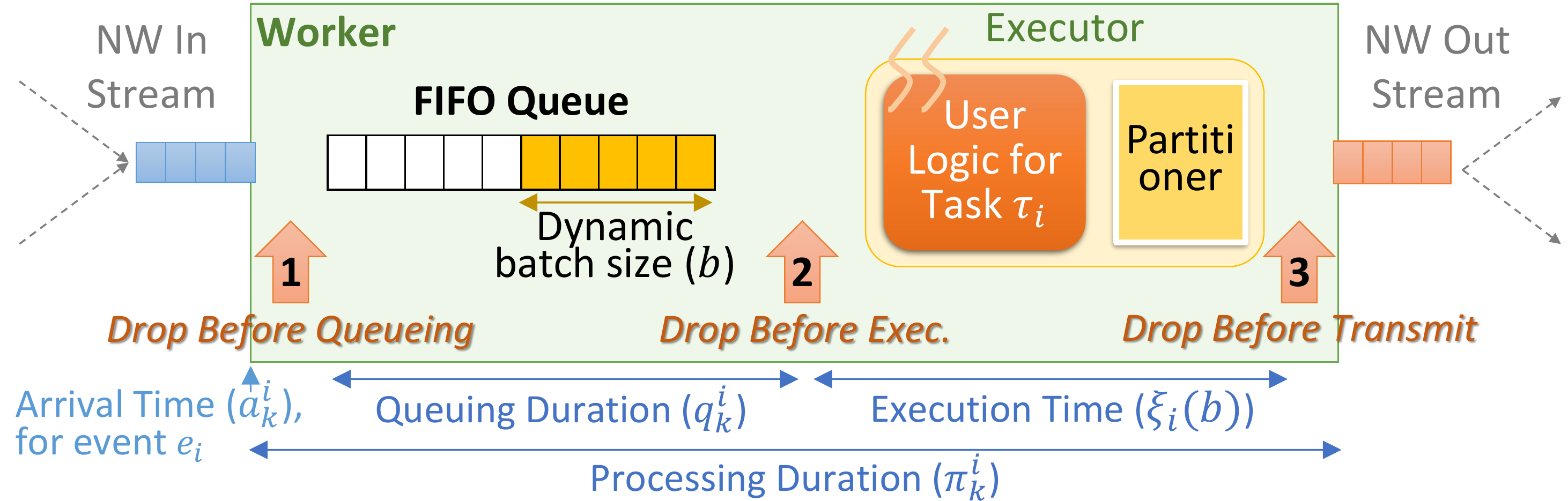}
	\caption{Processing events at a task, with batching and drops}
	\label{fig:task}
\end{figure}

The platform should drop any event $e_x^y$ that cannot reach the last task $\tau_n$ before a time $a^1_x + \gamma$ as it exceeds it maximum tolerable latency $\gamma$ and is hence \emph{stale}. So a task $\tau_i$ may drop an arriving event $e_x^i$ if $a^i_x > a^1_x + \gamma$. While simple, this waits till the allowed latency is exceeded and does not prevent resource wastage due to execution of tasks prior to the one where the event is dropped. E.g., if at tasks $\tau_{n-2}$ and $\tau_{n-1}$, we have $a^{n-2}_x < a^1_x + \gamma$ and  $a^{n-1}_x > a^1_x + \gamma$, then every event will be processed through the first $(n-2)$ tasks and yet dropped at the $(n-1)^{th}$ task, assuming that the task processing times and network performance stay constant. Ideally, the first task $\tau_1$ should reject a newly arriving event if it \emph{will} be rejected downstream to avoid resource wastage. 

We capture the potential staleness of an event at a task $\tau_j$ using a \emph{completion budget} $\beta_j$. This is the duration allowed for an arriving event to complete processing at this task, including the upstream time spent since its source task, i.e., if $u^i_k + \pi^i_k > \beta_i$ for an event $e^i_k$,  
it is stale and can be dropped. Since $\pi^i_k$ is not known when the event arrives but only after it is queued and executed, this drop decision is taken thrice within a task, as shown in Fig.~\ref{fig:task} and described below. 

This completion budget for a task can change often during the lifetime of an application as the system reacts to variability. Later, in \S~\ref{sec:runtime:budget}, we discuss how $\beta_i$ is actively updated to encapsulate this variability. It guarantees that for a given budget, if the downstream tasks do not exhibit further variability, then any event that meets the budget will be processed within $\gamma$, and vice versa.

\subsubsection{Drop Point 1} The first drop decision is when an event arrives at a task but before it is placed in its input queue (Fig.~\ref{fig:task}). This checks if the observed upstream time already expended plus the fastest possible execution duration for the event on this task, i.e., using a batch size of $b=1$, will cause the event to exceed it completion budget, even in the absence of any queuing. Since we do not know the actual queuing delay and batch size for this event at this time, we are conservative in this decision. So events that pass this test may still be dropped at subsequent drop points based on how long they spent in the queue and the actual execution duration. 
\hrule
\begin{algorithmic}[1]
	\small
	\Procedure{DropBeforeQueuing}{$a^1_k, a^i_k$} 
	\State $u^i_k = a^i_k - a^1_k$
	\If{$\big(u^i_k + \xi_i(1)\big) \leq \beta_i$} \textbf{~return} $false$ \Comment{\emph{Retain}}
	\Else \textbf{~~return} $true$ \Comment{\emph{Drop this event}}
	\EndIf		 
	\EndProcedure	
\end{algorithmic}
\hrule

\subsubsection{Drop Point 2} The second drop point is after the event is queued and put in a batch, but before the batch is executed. At this time, we have a batch of events $B$ of size $b$, which gives us the expected execution time $\xi_i(b)$, and the queuing duration $q^i_k$ for each of its events. If the predicted time to complete executing this event exceeds the completion budget, i.e., $u^i_k + q^i_k + \xi_i(b) > \beta_i$, we drop this event. The function is passed the entire batch and it returns an updated batch $B'$ without events that should be dropped. 
\vspace{0.05in}\hrule
\begin{algorithmic}[1]
	\small
	\Procedure{DropBeforeExec}{$B[~], b$} % t_2, a^1_k, B  
	\For {$\langle a^1_k, a^i_k, q^i_k, e^i_k \rangle$ in $B$}
	\State $u^i_k = a^i_k - a^1_k$
	\If{$\big(u^i_k + q^i_k  + \xi_i(b)\big) \le \beta_i$}  $B' \gets e^i_k$ \Comment{\emph{Retain}} %t_2 - a^1_k + \xi_i(b)
	\EndIf
	\EndFor
	\State \textbf{return} $B'$ \Comment{\emph{Events that should be executed}}
	\EndProcedure			
\end{algorithmic}
\hrule\vspace{0.05in}

\subsubsection{Drop Point 3} It is possible that the actual execution time was longer than estimated. So we trigger the third drop point after the batch execution, where the processing time $\pi^i_k$ has been spent on an event, but before its output events are sent on the output stream. Here, we check if the generated event $e^{i+1}_k$ at time $u^i_k + \pi^i_k$ has exceeded its completion budget $\beta_i$. This drop point is also important if the dataflow has branches, as discussed next. 
\hrule
\begin{algorithmic}[1]
	\small
	\Procedure{DropBeforeTransmit}{$a^1_k, a^i_k, \pi^i_k$} %t_3, a^1_k
	\State $u^i_k = a^i_k - a^1_k$
	\If{$( u^i_k + \pi^i_k) \leq \beta_i$} \textbf{~return} $false$ \Comment{\emph{Retain}}
	\Else \textbf{~~return} $true$  \Comment{\emph{Drop this event}}
	\EndIf		 
	\EndProcedure			
\end{algorithmic}
\hrule

By providing these three light-weight drop points, we achieve fine-grained control in avoiding wasted network or compute resources, and yet perform event drops just-in-time when they are guaranteed to exceed the budget. This balances application accuracy and performance. As a further optimization, we allow the user-logic to flag an event as \emph{avoid drop}, e.g., if it has a positive match, and the platform avoids dropping such events even if they exceed the tolerable latency. This can improve the accuracy and manage the active set size.

\subsubsection{Non-linear Pipelines}
While the drop logic has been defined for a linear pipeline, a module instance (task) in our dataflow can send an event to one of several downstream module instances, based on the partitioning function. However, the destination task for an output event is known only after the partitioner operates on that event, at drop point 3. The completion budget for a task depends on the network and compute performance of the downstream tasks that the event flows through, which can vary for the different task-paths taken. So for each task, we maintain one budget per downstream task. 

\subsection{Strategies for Dynamic Batching of Events}
\label{sec:runtime:batch}

Batching and executing events in a stream improves the throughput and reduces the average event latency~\cite{DBLP:journals/corr/CanzianiPC16}. When events arrive early at a task $\tau_i$ and/or the application has a relaxed $\gamma$, there may adequate completion budget $\beta_i$ to accumulate events from the input queue into a batch and execute them together, while not violating the budget and causing a drop. Since $\beta_i$ and the input event rates can vary over time, this batch size has to be dynamically decided.

We define the \emph{event deadline} $\delta^i_k = \beta_i + a^1_k$ for an event $e_k^i$ as the time at the task $\tau_i$ by which it must complete processing to avoid being dropped. Similarly, we define the \emph{batch deadline} $\Delta^i_p =  \min(\delta^i_1,...,\delta^i_m)$ as the latest time by which the batch $B_p$ having $m$ events must complete execution, and it is defined as the earliest event deadline among all events in the batch. Since temporal event ordering is not assumed, this may not be the first event in the batch.

The batching logic considers the event $e_x^i$ at the head of the queue at the present time $t_i$ for adding to the ``current batch'' $B_p$ having size $m$ by checking if $t_i + \xi_i(m + 1) > \min(\Delta^i_p, \delta^i_x)$, i.e., will adding this event to the batch cause the new execution time of the batch $(t_i + \xi_i(m + 1))$ to exceed the deadline of the batch $\Delta^i_p$ or the new event $\delta^i_x$. If not, we add the event to the current batch and update the batch deadline. We incrementally check and add events from the queue into the current batch. If the event at the head of the queue cannot be added to the batch, we submit the current batch for execution and add the head event to a new empty batch that becomes the current batch. 
Even if the queue is empty, the current batch is automatically submitted for execution when the local clock reaches the time, $\Delta^i_p - \xi_i(m)$. 

\subsection{Updating the Completion Budget}
\label{sec:runtime:budget}

The \emph{completion budget} $\beta$ for a task is central to determining the events to be dropped as well as the batch size. To deal with the dynamism in the system, the budget for all tasks must change over time. To enable this, each task $\tau_i$ stores a 3-tuple $\langle d^i_k, q^i_k, m^i_k \rangle$ for every event $e_k^i$ it has processed: the \emph{departure time} $d^i_k = u^i_k + \pi^i_k$, which sums the upstream time and the processing duration; the \emph{queuing duration} $q^i_k$; and the \emph{batch size} $m^i_k$ that the event was part of. Further, each downstream event sent by task $\tau_i$ in the pipeline is augmented with two header fields: the \emph{sum of execution times} $\overline{\xi}^i_k = \sum_{j=1..i}\xi_j(m^j_k)$ and the \emph{sum of the queuing delay} $\overline{q}^i_k = \sum_{j=1..i}q^j_k$, spent at the preceding tasks.

As an event executes through the pipeline, we either increase or decrease the budgets for the upstream tasks based on whether the event arrives at the destination task early or is dropped by a task in-between, respectively. The logic used for these budget changes are described next.

\subsubsection{Reducing the budget} If an event is processed within its completion budget at a task, it should also complete processing that pipeline within the maximum tolerable latency, if there is no downstream variability. However, if an event $e_k$ gets dropped at task $\tau_j$, it means that the downstream latency has deteriorated and hence, the completion budget of all the upstream tasks $\{ \tau_i | i=1..j-1\}$ must be reduced. If the event  has exceeded the completion budget by $\epsilon=d^i_k - \beta_i$, then the sum of the upstream completion budgets must be reduced by $\epsilon$. Intuitively, we reduce the budget at each upstream task $\tau_i$ proportional to the time spent in the queue and batch before execution. %ratio of batching delay at that task to the total upstream batching delay.
This causes batches with fewer events to be formed for execution. Using just the queuing time ratio for reducing the budget also avoids penalizing tasks with longer execution times.

Let $\overleftarrow{\lambda}^i_k$ be the duration by which the budget $\beta_i$ at an upstream task $\tau_i$ has to be \emph{reduced} due to an event $e^j_k$, being dropped at $\tau_j$, where $i<j$.
\begin{center}
	$\overleftarrow{\lambda}^i_k = \min(\epsilon \times  \frac{q^i_k}{\overline{q}^j_k},~~ \xi_i(m^i_k) - \xi_i(1))$
\end{center}

\noindent The first term in the $\min$ operator reflects the excess time $\epsilon$ scaled by the ratio of the queuing delay for the task relative to the sum of the delays at all the tasks upstream of the dropping task. The second term ensures that the budget reduction does not fall below the minimum possible budget required when streaming the event through with $b=1$. \\

Whenever an event is dropped at $\tau_i$, it sends a \emph{reject signal} to its upstream tasks with the event ID $k$, the excess duration over the budget $\epsilon$  and the sum of the queuing delays $\overline{q}^j_k$. The receiving task $\tau_i$ combines these with the 3-tuple it maintains for the event to calculate $\overleftarrow{\lambda}^i_k$, and updates its budget as:
\begin{center}
	$\beta_i^{new} = \min(d^i_k - \overleftarrow{\lambda}^i_k, ~~\beta^{old}_i)$
\end{center}
\noindent The first term determines the updated budget as the earlier departure time for that event, less the reduction in budget. Here, the $\min$ operator selects the lower of the previous and the new budget to make the model be resilient to out of order accept or reject signals.

\subsubsection{Increasing the Budget}
Events that arrive at the final task much earlier than the maximum tolerable latency indicate lost opportunity costs in improving the throughput and scalability of the pipeline by forming larger batches. Therefore, when an event arrives at the final task at $\epsilon=\beta_n-u^i_k$ duration earlier than its completion budget $\beta_n=\gamma$, and this value is greater than some set threshold, $\epsilon^{max}$, the completion budget of the upstream tasks must be increased. Intuitively, we increase the budget of a task proportional to its execution time, relative to the total execution times for all upstream tasks. This gives more weight to tasks with longer execution times, allowing them to increase their throughput which is likely to be the least in the pipeline.

If $\overrightarrow{\lambda}^i_k$ is the duration by which the budget $\beta_i$ at an upstream task $\tau_i$ has to be reduced due to an event $e^n_k$ completing ahead of time at the final task $\tau_n$, then:
\begin{center}
	$\overrightarrow{\lambda}^i_k = \min(\epsilon \times \frac{\xi_j(m^i_k)}{\overline{\xi}^{n-1}_k}, ~~ (m^{max} - m^i_k)  \times\frac{q^i_k}{m^i_k} + \xi_i(m^{max}) - \xi_i(m^i_k))$ 
\end{center}

\noindent The first term in the $\min$ operator scales the $\epsilon$ by the relative time spent in the execution duration for the task $\tau_i$, relative to the execution time at all tasks until (but not including) the final task. The second term ensures that the budget does not exceed the time taken to create and execute the largest batch size $m^{max}$ allowed by the user. % user defined \emph{maximum batch size} $m_{max}$. 
This assumes that the queuing time scales linearly with the number of events. As the prior budget already considers the queuing and execution time for a batch size $m^i_k$, we subtract it from $m^{max}$. 

A batch will have events with different queuing duration but the same batch execution duration. So some events in the batch will always arrive at the final task before $\gamma$ elapses. However, we should not increase the budget based on these early events in a batch. Rather, the decision to increase the budget is made only if event with the highest latency in a batch is below $\gamma+\epsilon^{max}$. If so, the task $\tau_n$ sends an \emph{accept signal} to all the upstream tasks with the slowest event's ID $k$, the duration of early arrival $\epsilon$ and the sum of upstream execution time, $\overline{\xi}^{n-1}_k$. These are used to calculate the value of $\overrightarrow{\lambda}^i_k$ at tasks $\tau_i$ and update their budgets using the 3-tuples for that event:
The completion budget for an task $\tau_j$ is increases as follows:
\vspace{-0.05in}\begin{center}
	$\beta_i = \max(d^i_k + \overrightarrow{\lambda}^i_k,~~\beta^{old}_i)$
\end{center}\vspace{-0.05in}
\noindent As before, selecting the $\max$ is against the previous budget is to make the model resilient to out of order signals.

The task budgets are increased when an event successfully reaches the final task ahead of time. But transient conditions may cause the system to reduce the budgets to such a low value that no subsequent events flow through to the final task without being dropped. In such cases, even if the conditions improve, the budget may never get updated. To address this, the system periodically sends \emph{probe signals} for every $k^{th}$ event that is dropped at a task $\tau_j$. This probe is forwarded downstream without being dropped. If this signal reaches the final task within $\gamma$, then the system calculates and sends the \emph{accept signal} so that the budget for the upstream tasks can be increased and regular events may start flowing through.

When \emph{bootstrapping} the application initially, the batch size for all tasks is fixed at $b=1$ and no budgets are assigned except $\beta_n=\gamma + a^1_k$. Subsequently, when accept or reject signals are triggered, these values are updated (without considering $\beta^{old}$) and they stabilize to the new budget.

\subsection{Formal Bounds and Time Synchronization} 
While our batching is not based on a fixed batch size but rather adapts to the events that arrive, we can formally bound the batch size and drop rate under certain assumptions. Later, we relax some of these assumptions.

\subsubsection{Fixed conditions}
We first derive bounds on the batch size and drop rate working under the assumption that a dataflow has constant known input rate $\omega$, 1:1 selectivity, no pipelining, the execution time matches the expectation, and the network and compute conditions are static. For simplicity during analysis, we also assume temporal ordering of events and no pipelining of the FIFO queue with the execution.

\para{Batch Size} Here, the goal is to estimate the bounds for the \emph{batch size} $m_i$ at a task $\tau_i$ when it has access to the budget and other variables at runtime, under a stable state. The inter-arrival rate between successive events can be written as $a^i_k - a^i_{k+1} = \frac{1}{\omega} \quad \forall k \in \mathbb{N}, i \in n$. Also, due to the temporal ordering of events, the batch deadline is bound by its first event.
Then, $m_i$ is the largest Integer value such that:
\vspace{-0.05in}\begin{center}
	$(m_i-1)\times\frac{1}{\omega} + \xi_i(m) \leq \beta_i - u^i_1 \qquad \text{ and } \qquad \xi_i(m_i) \leq \frac{\beta_i - u^i_1}{2}$
\end{center}\vspace{-0.05in}
\noindent In the first equation, we capture the intuition that the time to queue up the batch and to execute it~\footnote{There is no queuing delay for the first event, hence $(m-1)\times\frac{1}{\omega}$. For simplicity, we assume that the estimated execution time $\xi_i$ equals the actual execution time.} must not exceed the batch processing deadline. The second equation ensures the stability of the system such that the time to execute a batch does not exceed the deadline for the next batch being accumulated, i.e., execution time for the current batch should be less than the queuing time for next batch. Here, $\omega$ and $\xi_i(m)$ are unconstrained natural numbers while $\beta_i$ and $u_i$ are available at runtime. A solution for $m_i$ may not exist within these constraints, which means that the input rate $\omega$ is unsustainable. In such cases, events should be dropped.  

\para{Drop Rate} If no solution for $m_i$ exists above, then we find the \emph{drop rate} of events, $(\omega - \omega^{max})$, relative to the largest stable input rate, $\omega^{max}$, that can be support, and the associated batch size. The goal is to maximize $m_i$ and $\omega^{max}$ such that: 
\vspace{-0.05in}\begin{center}
	$(m_i-1)\times\frac{1}{\omega^{max}} + \xi_i(m_i) \leq \beta_i - u^i_1 \qquad \text{ and }$
\end{center}\vspace{-0.05in}
\vspace{-0.05in}\begin{center}
	$\xi_i(m_i) \leq \frac{\beta_i - u^i_1}{2}$
\end{center}\vspace{-0.05in}
Compared to streaming execution with $m=1$, batching adds latency to the overall event processing time while increasing the throughput. We quantify the %added 
increase in the \emph{average latency} per event caused by batching, relative to streaming, for a task $\tau_i$ as:
\vspace{-0.05in}\begin{center}
	$\frac{m_i-1}{2 \times \omega } + \xi_i(m_i) -\xi_i(1)$
\end{center}\vspace{-0.05in}
\noindent Here, the first term is the average queuing time for the $m_i$ events in the batch and the latter indicate the execution time for the batch.

\subsubsection{Resilience to Unsynchronized Clocks}
Devices in a WAN may have unmanaged devices that do not have tightly-synchronized clocks~\cite{buchholz2007brief}. While our drop and batch decisions are based on the timestamps at the different devices, these have been designed to withstand skews across the device clocks.
The drop logic as well as the completion budget are based on the \emph{upstream time}, with the other time variables within a task being defined relative to it. So making the upstream time resilient to unsynchronized clocks with consequently address all other time calculations.

Let $\kappa_1..\kappa_n$ be the clocks for the $n$ devices hosting the tasks in a pipeline. As stated before, we require that $\kappa_1 = \kappa_n$. Let the signed-values $\sigma_2..\sigma_{n-1}$ be the skew between the clocks $\kappa_2..\kappa_{n-1}$ relative to $\kappa_1$ and $\kappa_n$, i.e., $\sigma_i = \kappa_i - \kappa_1$.
The upstream time $u^i_k = a^i_k - a^1_k$ for an event $e_k^i$ arriving at a task $\tau_i$. When corrected for the skew, we have $\widetilde{u}^i_k = (a^i_k - \sigma_i) - a^1_k$ 

Similarly the update rule for the \emph{completion budget} $\widetilde{\beta}_i$ for task $\tau_i$, when corrected for skew, can be written by correcting the \emph{departure time} as $(\widetilde{d}^i_k - \overleftarrow{\lambda}^i_k)$ or $(\widetilde{d}^i_k + \overrightarrow{\lambda}^i_k)$, as is the case when updating using a reject or an accept signal. Here, $\widetilde{d}^i_k = \widetilde{u}^i_k + \pi^i_k = u^i_k - \sigma_i + \pi^i_k$, since $\pi$ is a duration calculated locally within a single device. Also, $\lambda$, the budget change factor, depends on $\epsilon$ and other local durations, with $\widetilde{\epsilon} = \widetilde{d}^i_k - \widetilde{\beta}_i = \epsilon$. Hence $\widetilde{\beta}_i = \beta_i - \sigma_i$, and it can be shown that $\widetilde{\beta}^{old}_i = \beta^{old}_i - \sigma_i$ with an inductive argument.

As a result, in all three of our drop points, when replacing $u^i_k$ and $\beta_i$ with their skew-correct forms (in lines 3, 4 and 3, respectively), we have a $-\sigma_i$ term added symmetrically to both the left and the right sides of the comparisons, which cancel each other out. This shows that our drop logic is resilient to clock-skews.

For batching, we use the test $\big(t_i + \xi_i(m + 1) > \min(\Delta^i_p, \delta^i_x)\big)$ to decide if we should add an event $e^i_x$ arriving at task $\tau_i$ to a batch $B_p$. Here too we can correct the skew for the time points $\widetilde{t}_i = t_i - \sigma_i$ and $\widetilde{\delta}^i_x = (\widetilde{\beta}_i + a^1_k) = (\beta_i - \sigma_i + a^1_k) = (\delta^i_x - \sigma_i)$; similarly, $\widetilde{\Delta}^i_p = \Delta^i_p - \sigma_i$ as it derives from $\widetilde{\delta}$. As we see, the $-\sigma_i$ term is added to both the sides of the comparison and cancel each other out, indicating that the batch size is resilient to unsynchronized clocks as well.

\section{Experiments}
\label{sec:results}
We perform targeted and detailed experiments to evaluate the benefits of the domain-sensitive \emph{Tuning Triangle knobs} (Fig.~\ref{fig:dataflow}, inset) we offer: (1) a smarter tracking logic, (2) dynamic batching capability, and (3) multi-stage dropping strategies. We empirically demonstrate our proposition that these knobs help achieve \emph{two of the three qualitative metrics}: (1) end-to-end latency within the user-defined threshold $\gamma$, (2) scaling to a large number of the active cameras, and (3) the accuracy of the tracking.

\subsection{Setup}
\para{System Setup} 
We mimic the resource conditions of $96$ Raspberry Pi 3B edge devices on a local cluster, which has 1 \emph{head node} and 10 \emph{compute nodes}. The compute nodes each have an 8-core/16-hyperthreads @2.10 GHz Intel Xeon CPU E5-2620 v4 CPU and 64GB DDR4 RAM, while the head node has the same CPU in a dual socket configuration and 512GB RAM. Each Xeon CPU core perform comparable to a 4-core Pi 3B, as measured using the CoreMark benchmark. All the nodes have a 1Gbps network interface. The nodes run Centos v7.5 with Linux 3.10.0 kernel release, Java 1.8 and Python v3. The head node hosts a Kafka v2.11.0 pub-sub broker for routing input video streams while the compute nodes have Pytorch v1.0.1 and Tensorflow 1.2 installed.

\para{\anv Setup} We have two \anv worker processes on each compute node and the head node. The number of FC instances equals the number of cameras used in that experiment. In addition, we have 10 VA, 10 CR, 1 TL and 1 UV instances. The FC instances are scheduled across the $10$ compute nodes in a round-robin manner for load balancing, and run on one of the two workers on the node. The VA and CR instances are also placed in a round-robin manner on these nodes, on the other worker. This co-locates a subset of the FC, VA and CR on the same server and minimizes their network transfer overheads. Since each instance runs on a separate executor thread within the worker, each in-effect runs on a Pi 3B-class CPU core. The TL and UV instances run on a worker each on the head node. 

\para{Applications} We implement two tracking applications, \emph{App 1} and \emph{App2}, described in Table.~\ref{tab:applications} and evaluate them in our experiments. These omit the QF module given its nascency. Further, we use three TL algorithms for the applications. \emph{TL-Base} is a na\"{i}ve baseline that keeps all the cameras in the network active all the time. \emph{TL-BFS} has access to the underlying road network, but assumes a fixed road-length % of $84~m$ 
for all edges when performing the spotlight BFS strategy. \emph{TL-WBFS} is similar, but aware of the exact lengths of each road segment (Alg.~\ref{alg:app1}).
Both \emph{TL-BFS} and \emph{TL-WBFS} are configured with the \emph{expected peak speed} of the entity being tracked, which varies across experiments. The maximum tolerable latency is set as $\gamma=15~secs$. We provide a detailed analysis for App 1, and confirm similar for App 2 as well.

\para{Workload} For the road network, we extracted a circular region of $7~km^2$ from \emph{Open Street Maps}~\cite{OpenStreetMap}, centered at the Indian Institute of Science, Bangalore campus. This has $1,000$ vertices and $2,817$ edges, with an average road length of $84.5~m$. We use this as the fixed road length for TL-BFS. We use the \emph{CUHK03} Person Re-identification image dataset~\cite{li2014deepreid} with $1,360$ unique targets and $10,531$ images, which provide \emph{true positives or negatives} for the models used. Each JPG image is $64\times128 px$ in size with RGB colors, and a median size of $2.9~kB$. 

We use these images to simulate video feeds that mimic the movement of the query entity through a road network. The simulator takes as input the road network with the road lengths, the speed of the entity being tracked, their starting vertex in the network, and the labelled images for the entity. A given number of cameras are ``placed'' on vertices surrounding the starting vertex. The number of cameras is $1000$ by default, but varies for some experiments. We simulate the movement of the entity from the source vertex as a \emph{random walk} at a speed of $1~m/sec$ ($3.6~km/hr$). Each camera generates a timestamped feed of images at $1~fps$ using the true negative images (i.e., images not containing the entity), but uses the true positive images for the time intervals when the tracked entity is within the camera's FOV during the walk. For each camera, the simulator publishes its image feed in real-time to a unique topic using the Kafka broker. The FC module for the camera subscribes this appropriate topic to acquire the input stream.

\para{Baseline} We also design a \emph{near-optimal baseline (NOB)} to evaluate the effectiveness of our dynamic batching. This uses prior benchmarking on the stable system to determine the smallest batch size that can meet specific input rates without any drops or delays, for rates of $1$--$1000~events/sec$, in steps of $10$. This forms a lookup table. During the application execution, the platform dynamically picks the batch size for the rate closest to the current input rate from this table. Such a runtime strategy can maximize the throughput while also minimizing the latency. In fact, under static system conditions, this will be near-optimal, limited only by the discrete rates for which the table is constructed. 

\subsection{Analysis of App 1}

\subsubsection{Analysis of Batching Strategy}
\label{sec:exp:batching}

\begin{figure}[t]
	\centering
	\subfloat[\emph{es=4~m/sec}]{\label{fig:lat_violin}
		\includegraphics[width=0.47\columnwidth]{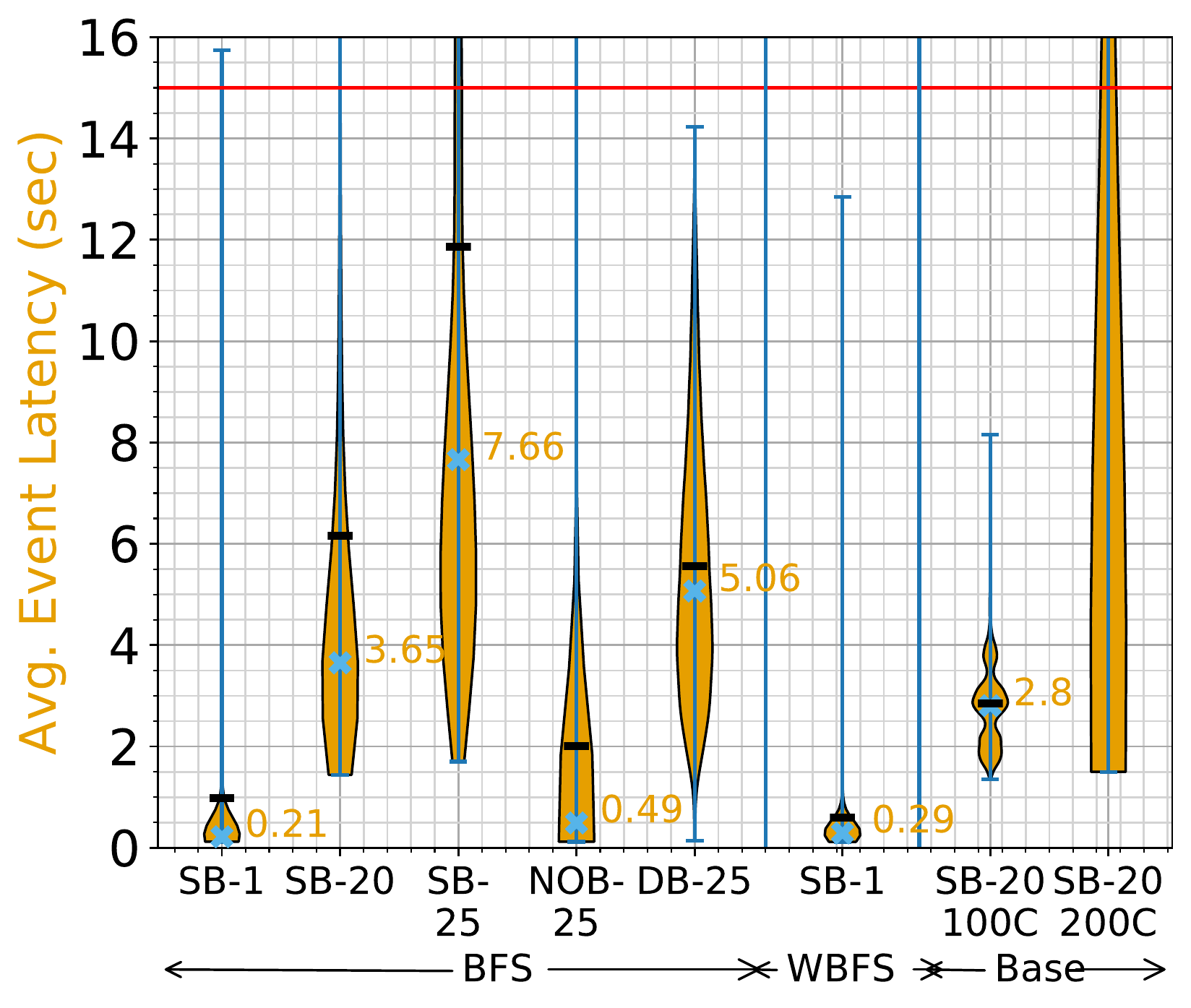} %CR_runtime
	}
	\subfloat[\emph{es=6~m/sec}]{\label{fig:lat_violin_6ms}
		\includegraphics[width=0.28\columnwidth]{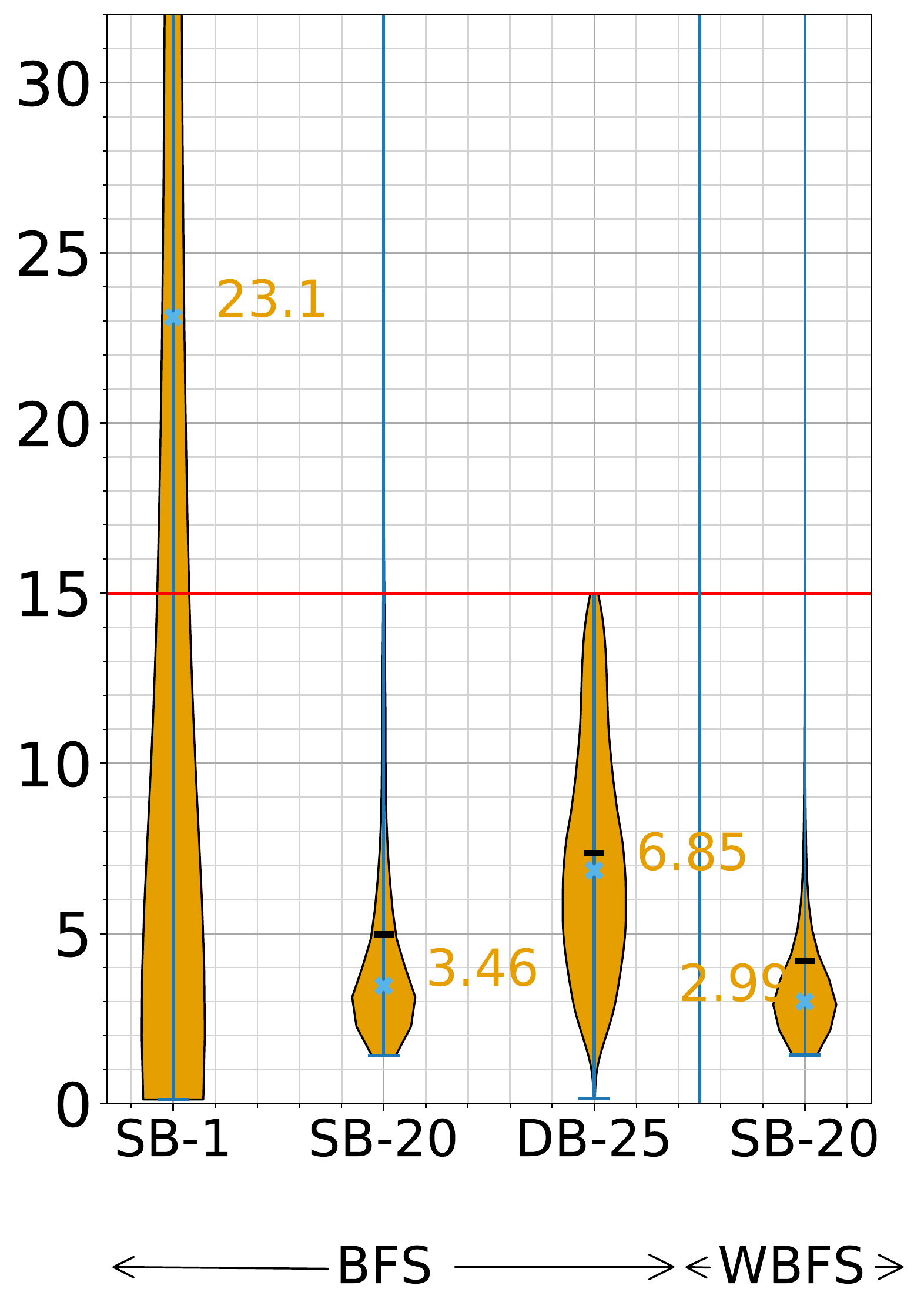}
	}
	\subfloat[\emph{es=7~m/sec}]{\label{fig:lat_violin_7ms}
		\includegraphics[width=0.22\columnwidth]{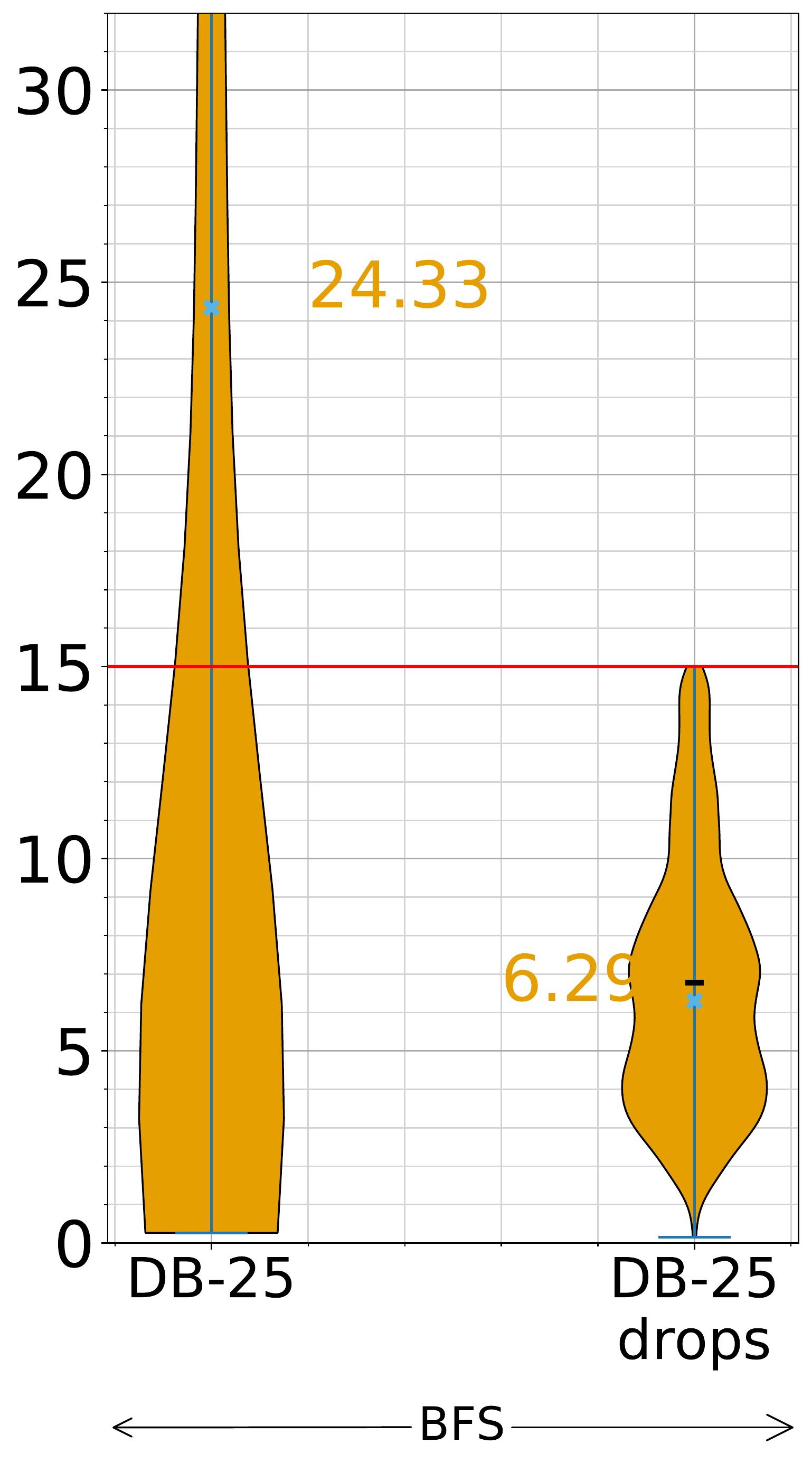}
	}
	\caption{Distribution of the average end-to-end event latencies for the different batching and TL strategies of App 1}
\end{figure}

\begin{figure}[t]
	\centering
	\subfloat[\emph{es=4~m/sec}]{\label{fig:bar_4ms}
		\includegraphics[width=0.47\columnwidth]{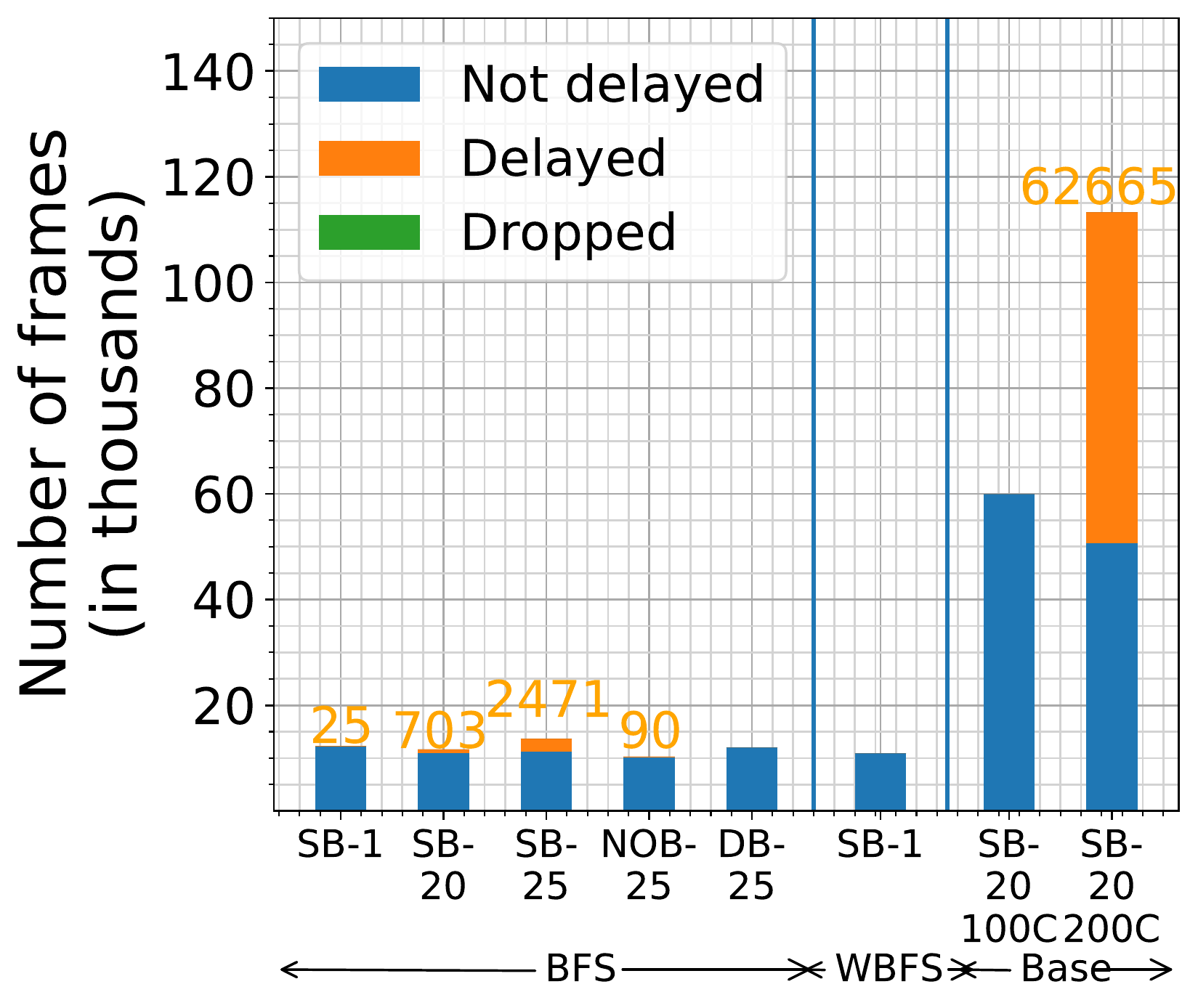} 
	}
	\subfloat[\emph{es=6~m/sec}]{\label{fig:bar_6ms}
		\includegraphics[width=0.28\columnwidth]{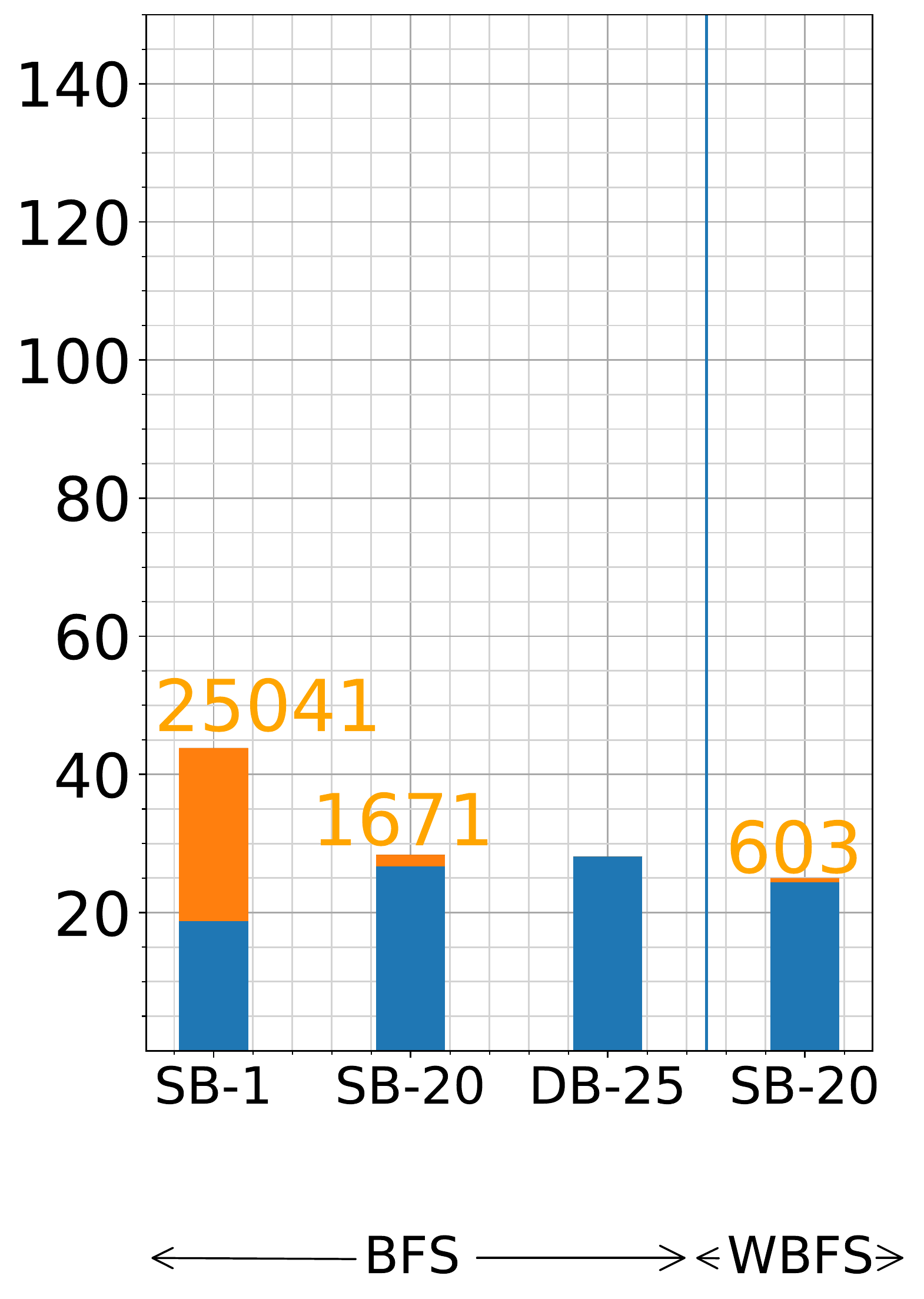}
	}
	\subfloat[\emph{es=7~m/sec}]{\label{fig:bar_7ms}
		\includegraphics[width=0.22\columnwidth]{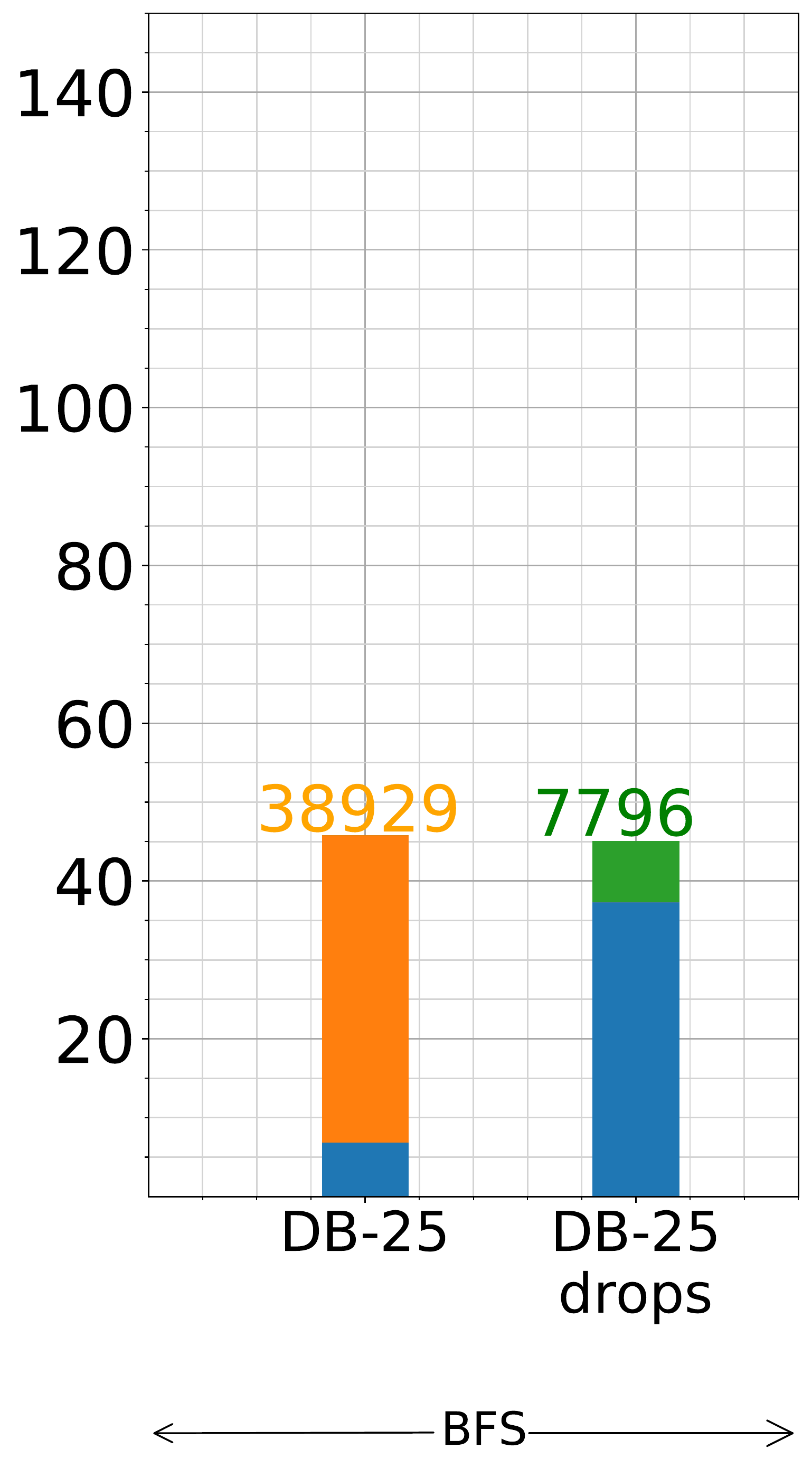}
	}
	\caption{Events with latency $\leq \gamma$ vs. Delayed vs. Dropped events, for different peak speeds, batching and TL strategies}
\end{figure}

We first examine the benefits of the \emph{dynamic batching} knob of the tuning triangle, while keeping the tracking logic fixed at \emph{TL-BFS} and \emph{disabling drops}. Further, the entity's \emph{peak speed} is configured as $es=4~m/sec$ with TL-BFS. This is set based on an estimated peak speed since underestimating can cause the entity to be lost due to a slow \emph{Rate of Expansion (RoE)} of the spotlight when the entity is in a blindspot; too high a value can cause a fast spotlight RoE and large active camera count that overwhelm the resources (or cause drops).

\begin{figure*}[t]
	\centering
	\subfloat[Streaming ($b=1$)]{\label{fig:streaming}
		\includegraphics[width=0.5\textwidth]{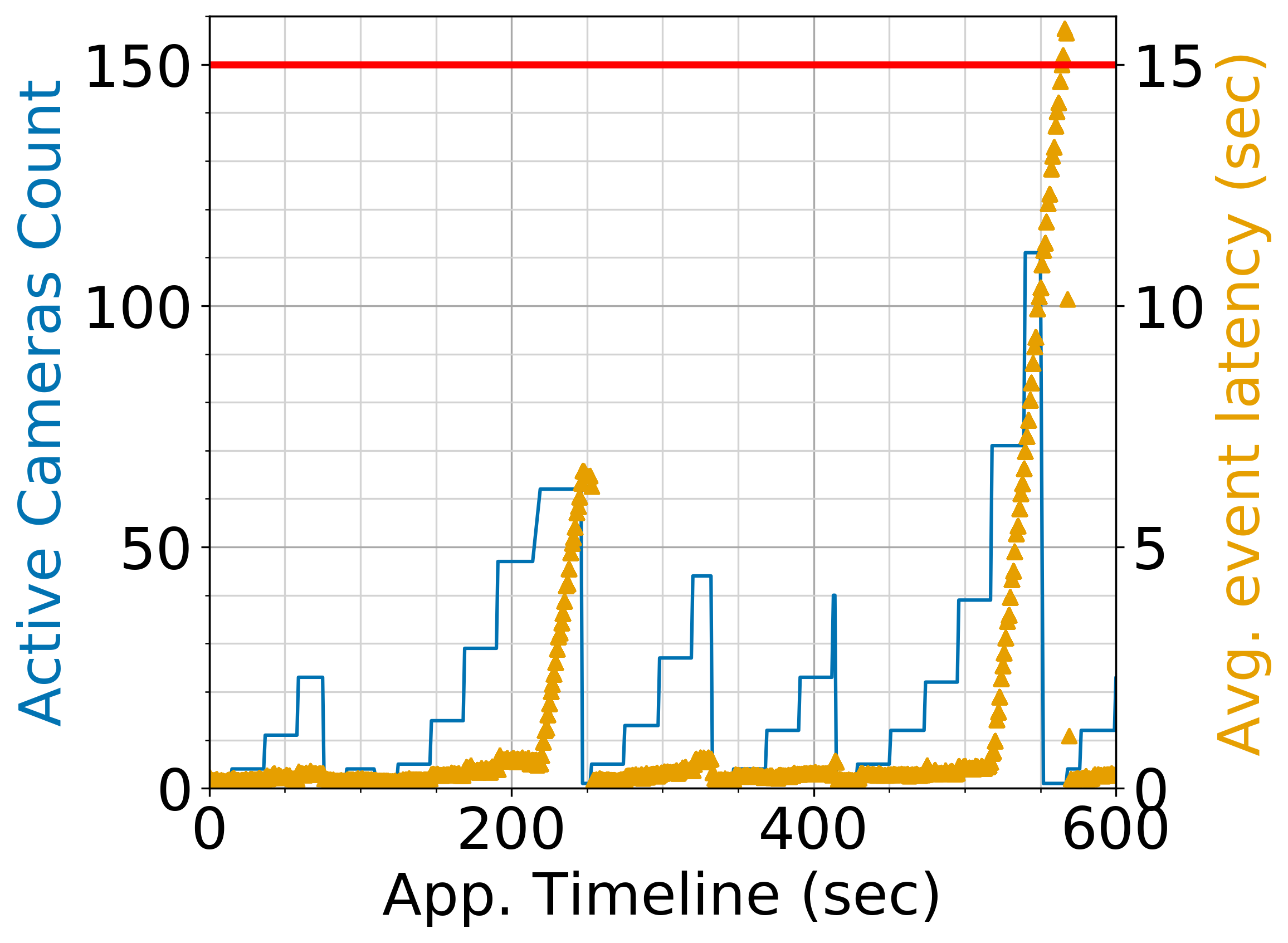}
	}
	\subfloat[Static Batching ($b$=$20$)]{\label{fig:4ms_fixed:bs:20}
		\includegraphics[width=0.5\textwidth]{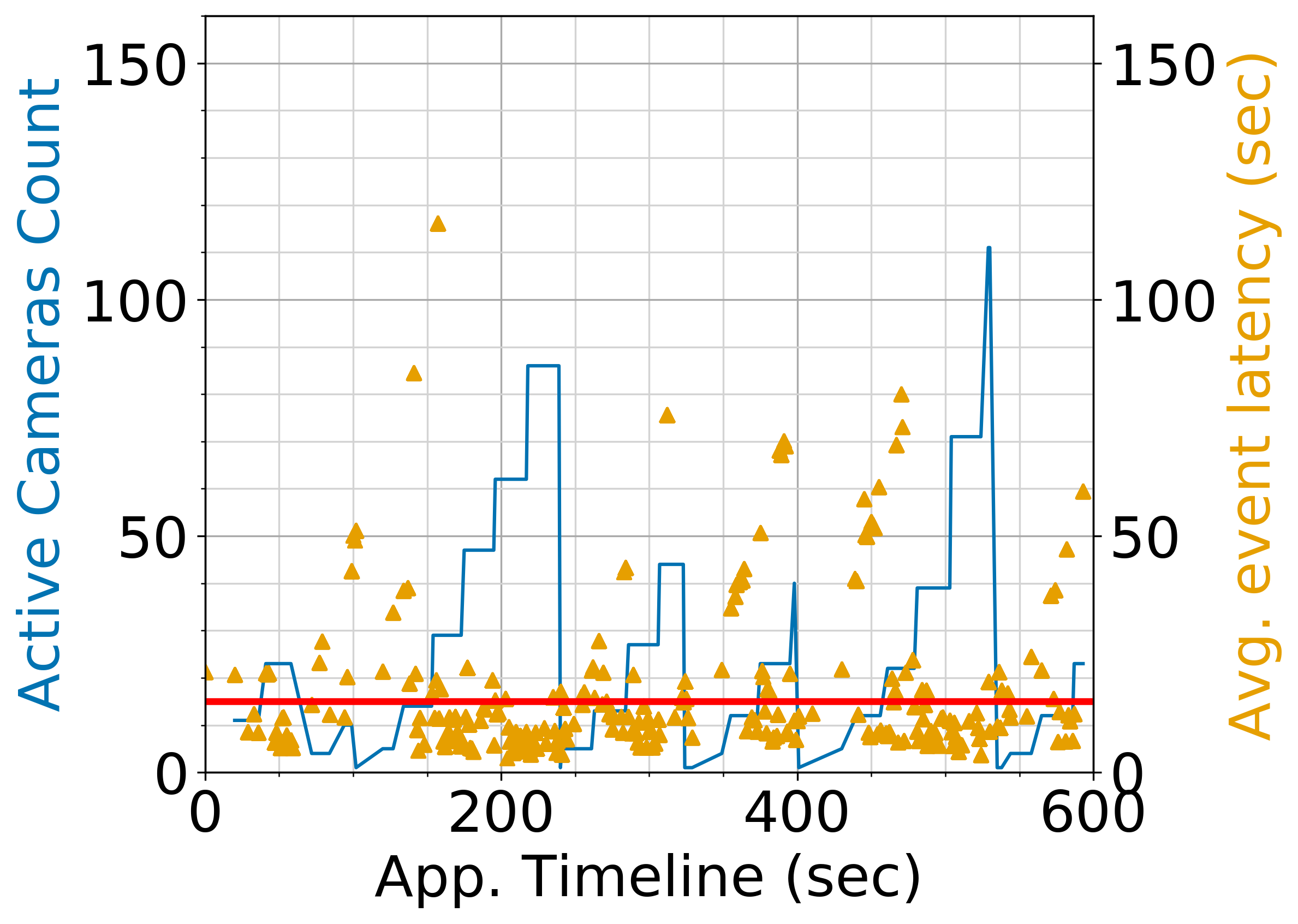}
	}\\
	\subfloat[NOB Batching ($b^{max}$=$25$)]{\label{fig:4ms_base_dynamic}
		\includegraphics[width=0.5\textwidth]{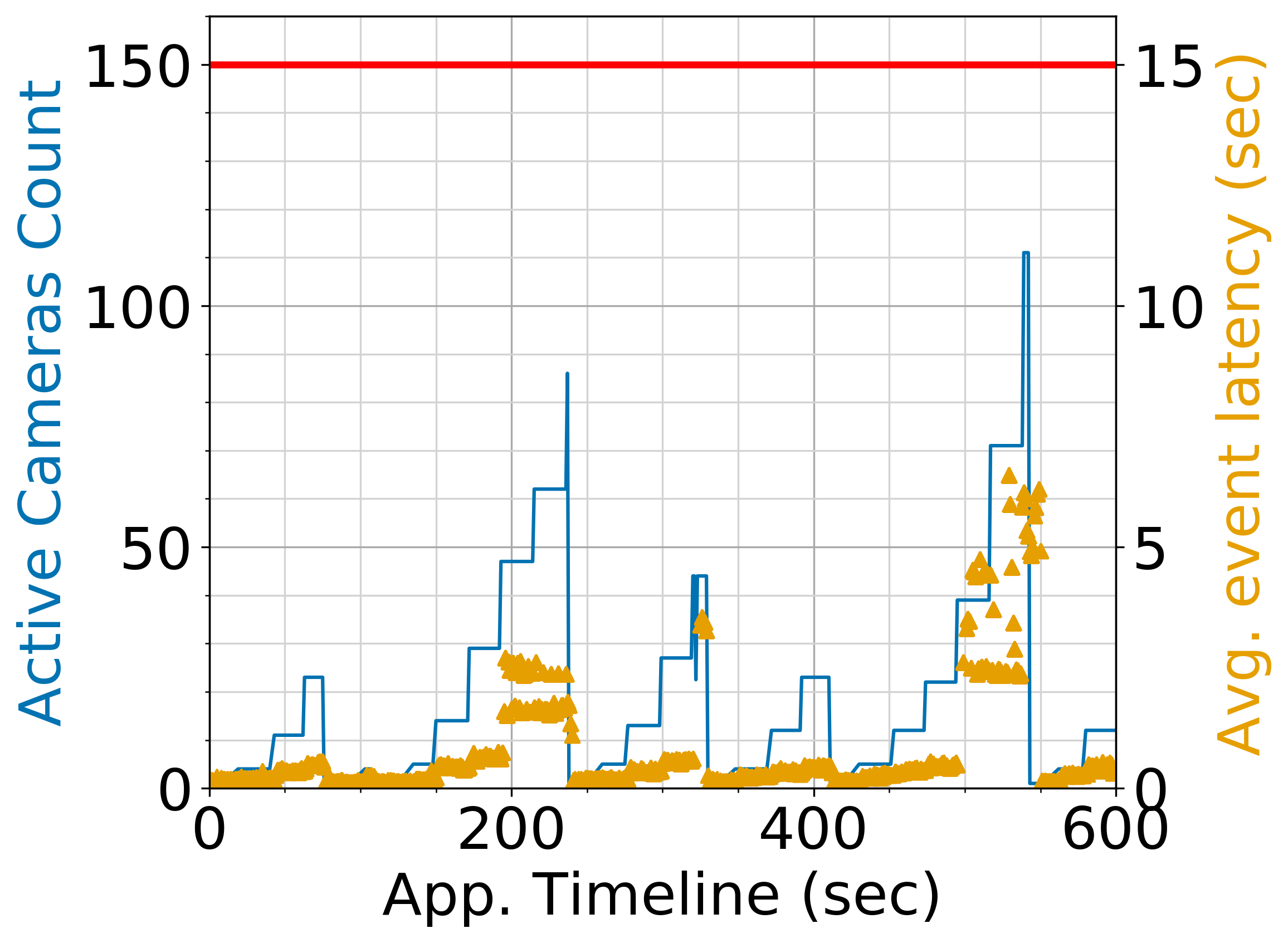}
	}
	\subfloat[\anv Batching ($b^{max}$=$25$)]{\label{fig:4ms_dynamic}
		\includegraphics[width=0.5\textwidth]{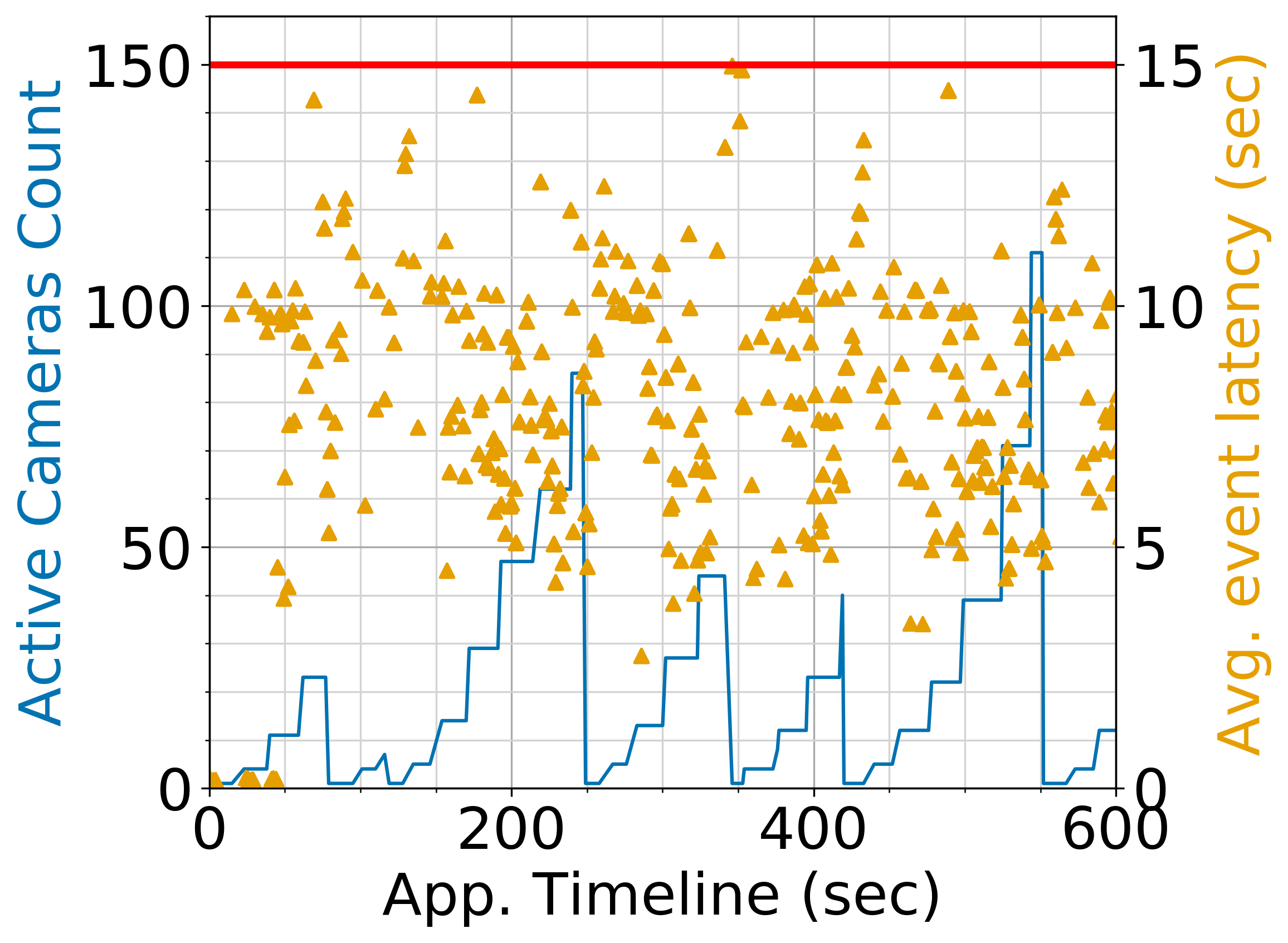}
	}
	\caption{\emph{\# of active cameras} (left Y axis, blue line) and \emph{Avg. end-to-end event latency} (right Y axis, yellow dots) over \emph{Application execution timeline} (X axis) for different batching strategies, TL-BFS, $es=4~m/sec$. Red horizontal line shows $\gamma=15~secs$.}
\end{figure*}

\para{Need for Batching} The \anv dataflow can be executed in a streaming manner without batching, i.e., a batch size of $b=1$. But this sacrifices the input throughput, and hence scalability of the number of active cameras that can be supported.

Fig.~\ref{fig:streaming} shows the application timeline plot for the streaming configuration, using a static batch of size $b=1$ (SB-1). Here, the X axis is the wall-clock time (secs) of the tracking application's execution, the blue line on the left Y axis is the number of active cameras as decided by TL, and the yellow dots on the right Y axis are the end-to-end latency of events (frames) from the source to the sink averaged over every second of the application's execution time. A red horizontal line shows the maximum tolerable latency, $\gamma=15~secs$. Further, Fig.~\ref{fig:lat_violin} shows a violin distribution of these $1~sec$ average event latencies for the streaming and other batching configurations. Similarly, Fig.~\ref{fig:bar_4ms} shows the number of events ($1000\times$) processed within the deadline of $\gamma$ against those that were delayed (orange, labelled).

In Fig.~\ref{fig:lat_violin}, streaming (SB-1) exhibits the lowest median latency, at $\approx 200 ms$, much below $\gamma=15~secs$ that is acceptable. However, this is at the cost of the latency occasionally exceeding the threshold, as seen from the violin outliers, and the $25$ delayed events in Fig.~\ref{fig:bar_4ms}. In fact, if we configure TL with an entity peak speed of $es=6~m/s$, $57\%$ of the input events exceed $\gamma$ when streaming (Fig.~\ref{fig:bar_6ms}).

As Fig.~\ref{fig:streaming} shows, these delayed frames occur when the active camera count exceeds 100. Here, the blue lines exhibit a saw-tooth behavior -- the spotlight logic increases the active set of cameras when the entity is in a blindspot, and this drops to $1$ when the entity is reacquired within the FOV of an active camera. At $\approx 550~secs$, the entity is in a blindspot long enough that the count spikes to $111$ cameras, stressing the available resources and causing the latency to grow to $16.8~secs$.

Specifically, this latency is caused by CR, whose DNN is the slowest task the dataflow at $120~ms/event$ and can service $\mu=\frac{1000}{120}=8.33~events/sec$ per CR instance. For an event arrival rate $\lambda > \mu$, the queue in unstable and the queuing delays will grow exponentially. At the $550^{th}$ second, the feeds from $111$ active cameras at $1~fps$ are mapped to $10$ CR instances, and $8$ of these CR instances receive events faster than $\mu$ causing a spike in the event latency. However, in this case, the system is able to recover at $\approx 570~secs$ when the active set size drops to $1$. 

In summary, it is vital to process \emph{all} the frames in a \emph{timely} manner rather than \emph{many} in a \emph{fast} manner, i.e., taking up to but below $\gamma$ is better. Batching offers such efficiencies and helps scale to a larger number of active cameras.

\para{Limitations of Static Batching}
While batching improves the system throughput, using a fixed or static batch size has two limitations. There can be a missed detection of the entity or certain frames can experience a very high end-to-end latency. Next, we provide empirical evidence for this.

We first set the static batch size $b=20$ for all tasks, and report its latency, active camera count and events processed without delay in Figs.~\ref{fig:4ms_fixed:bs:20},~\ref{fig:lat_violin} and~\ref{fig:bar_4ms}, as before. We see that the median latency has increased to $3.65~secs$ in Fig.~\ref{fig:lat_violin}, and there are no sharp growths in the latency in Fig.~\ref{fig:4ms_fixed:bs:20} which would indicate an unstable queue. Interestingly, in periods where the active camera count increases, like between $140$--$240~secs$, the mean latency decreases -- more cameras means a higher input rate, which fills up a batch and triggers it faster, thus reducing batching delays.

But better stability does not always result is fewer latency violations. In Fig.~\ref{fig:bar_4ms}, we see $6\%$ or $703$ events exceed $\gamma$, more than the streaming case. Since the batch size is fixed, \emph{there is no bound} on the time spent by an event waiting for the batch to be completed before execution. This causes events to be delayed.
However, if using $es=6~m/s$, none of the input events exceed $\gamma$ (Fig.~\ref{fig:bar_6ms}). This too indicates that a fixed value of the batch size will not suffice for all situations. Selecting an appropriate peak latency $\gamma$ and peak speed $es$ by the domain user is also important.

Another consequence of event delays due to a large fixed batch size is that TL grows the spotlight larger than necessary as it receives negative detections of the entity late, and also delays shrinking the spotlight as positive detections arrive late. These put further stress on the resources. E.g., at the $220^{th}$ second, the active set size is 62, 86 and 139 using a batch sizes of $b=1, 20$~and~$25$ (latter is not plotted), due to the increase in median latency from $0.2~sec$ to $3.65~secs$ and $7.6~sec$ (Fig.~\ref{fig:lat_violin}). When $b=25$, there is a delay of $22~secs$ in detecting a missing entity, due to which the active camera count sharply jumps to $139$, and also causes a detection of the entity in the neighboring cameras to be missed.
In fact, $22\%$ of events are delayed for $b=25$ (Fig.~\ref{fig:bar_4ms}).

\para{Benefits of Dynamic Batching}
The varying number of active cameras and its consequence on the latency is a strong motivation for using a dynamic batch size. Here, we compare \anv's dynamic batching strategy with the Near-optimal Baseline (NOB) dynamic batching strategy. The maximum batch size is limited to $b^{max}=25$ for both.

The timeline plot for NOB and our dynamic batching (DB-25) are shown in Figs.~\ref{fig:4ms_base_dynamic} and~\ref{fig:4ms_dynamic}, with their latency distribution and delay frames listed in Figs.~\ref{fig:lat_violin} and~\ref{fig:bar_4ms}. They key observation is that there are no delayed events in \anv's batching while $90$ events are delayed for NOB, at time periods $350~secs$ and $520~secs$. The latter is despite selecting a near-optimal batch size from the lookup table as the system is executing. Even minor runtime variations cause a larger batch size to be selected, leading to events being queued longer when forming the batch and delaying them. But \anv's batching prevents delays in \emph{all cases}.

NOB does offer a lower latency distribution, at a median of $0.4~secs$ (Fig.~\ref{fig:lat_violin}). The batch size it chooses is often between $2$ and $5$, thus approaching a streaming scenario. The median latency for \anv's batching is $7.66~secs$, with a wide variety of batch sizes. But as discussed before, reducing the latency is not a goal; just ensuring that all events reach within $\gamma$.

\begin{figure}[t]
	\centering
	\subfloat[VA Camera count \& Batch size]{\label{fig:4ms_dynamic_bs:va}
		\includegraphics[width=0.47\columnwidth]{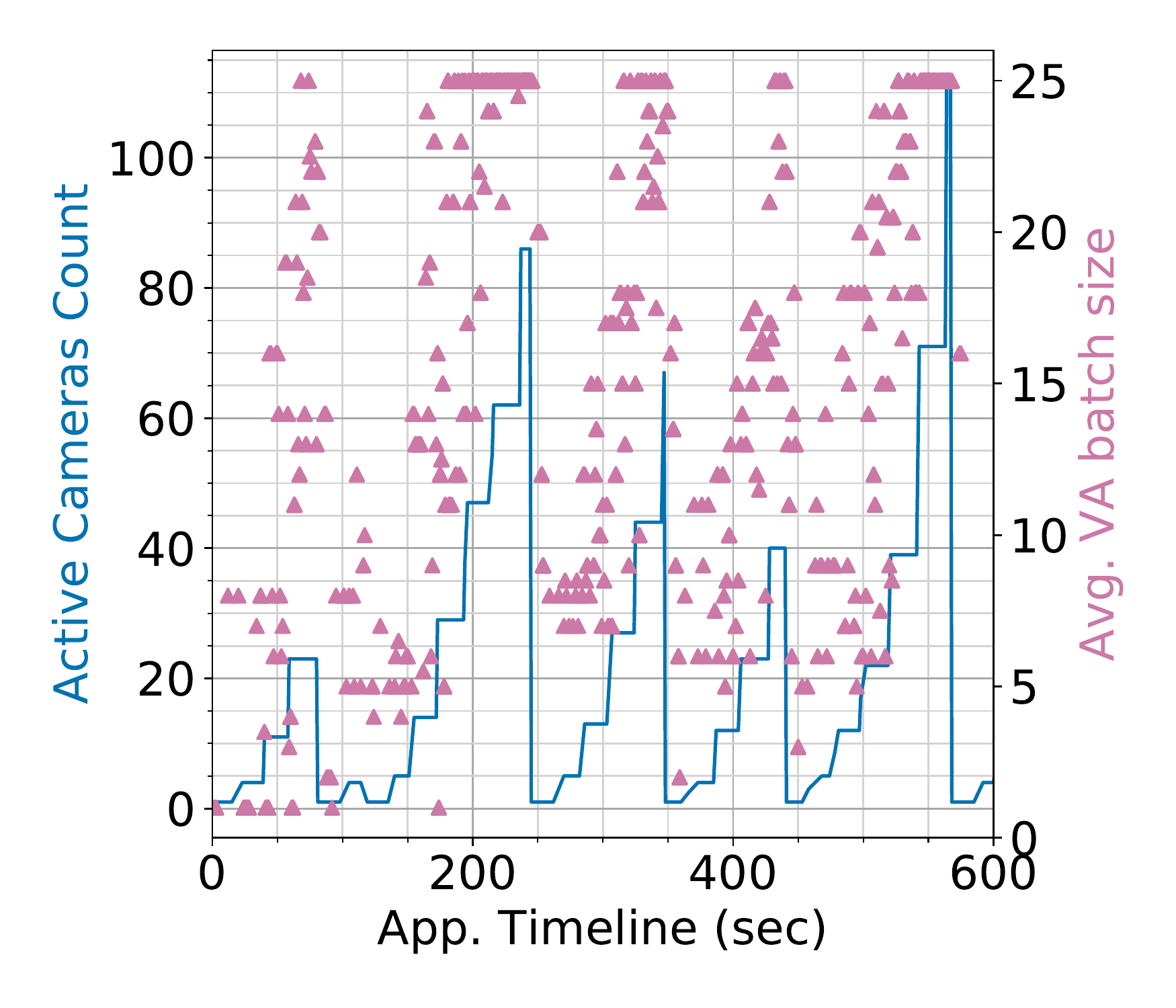}
	}
	\subfloat[CR Camera count \& Batch size]{\label{fig:4ms_dynamic_bs:cr}
		\includegraphics[width=0.47\columnwidth]{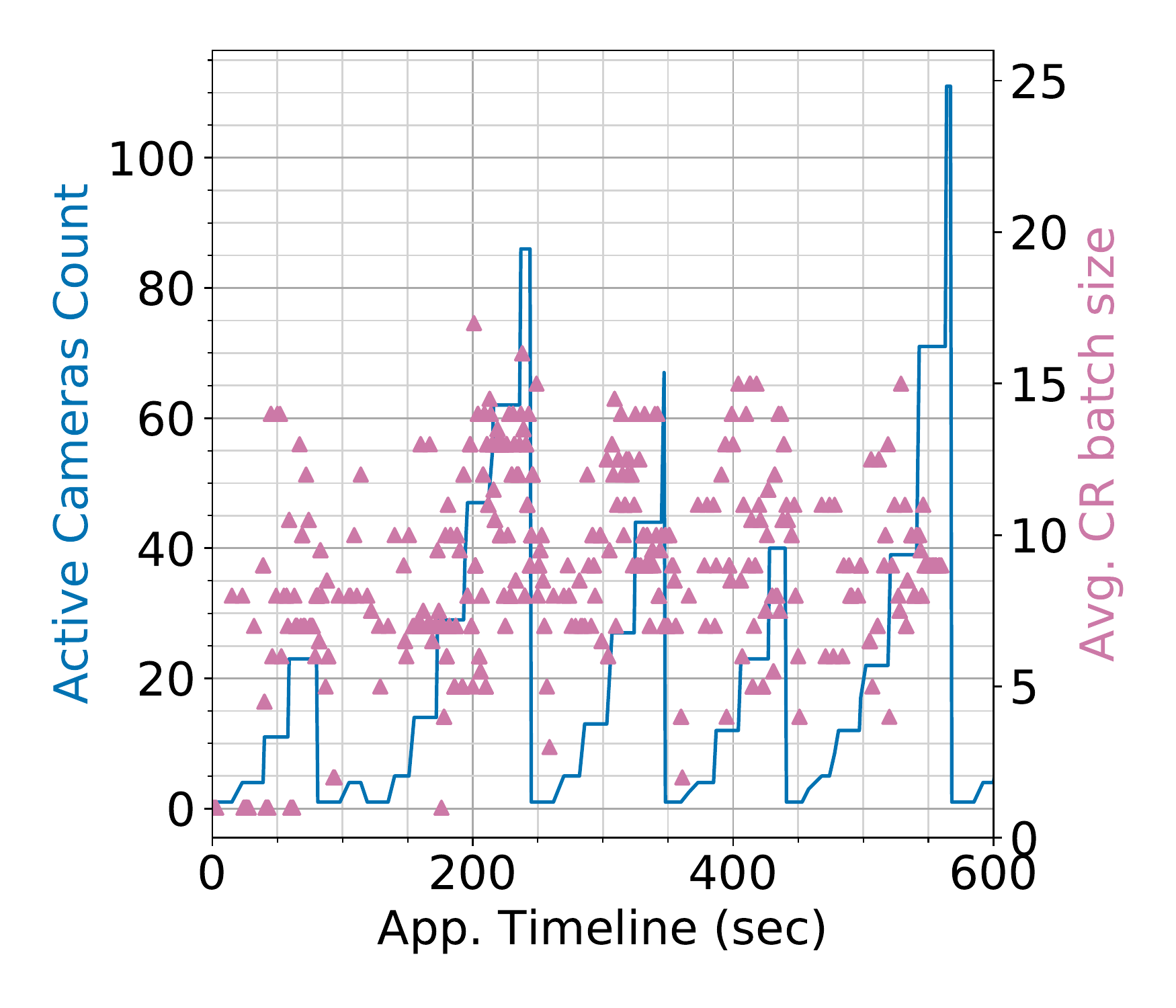}
	} \\
	\subfloat[VA Batch size vs. Event latency]{\label{fig:va_lat_4ms}
		\includegraphics[width=0.47\columnwidth]{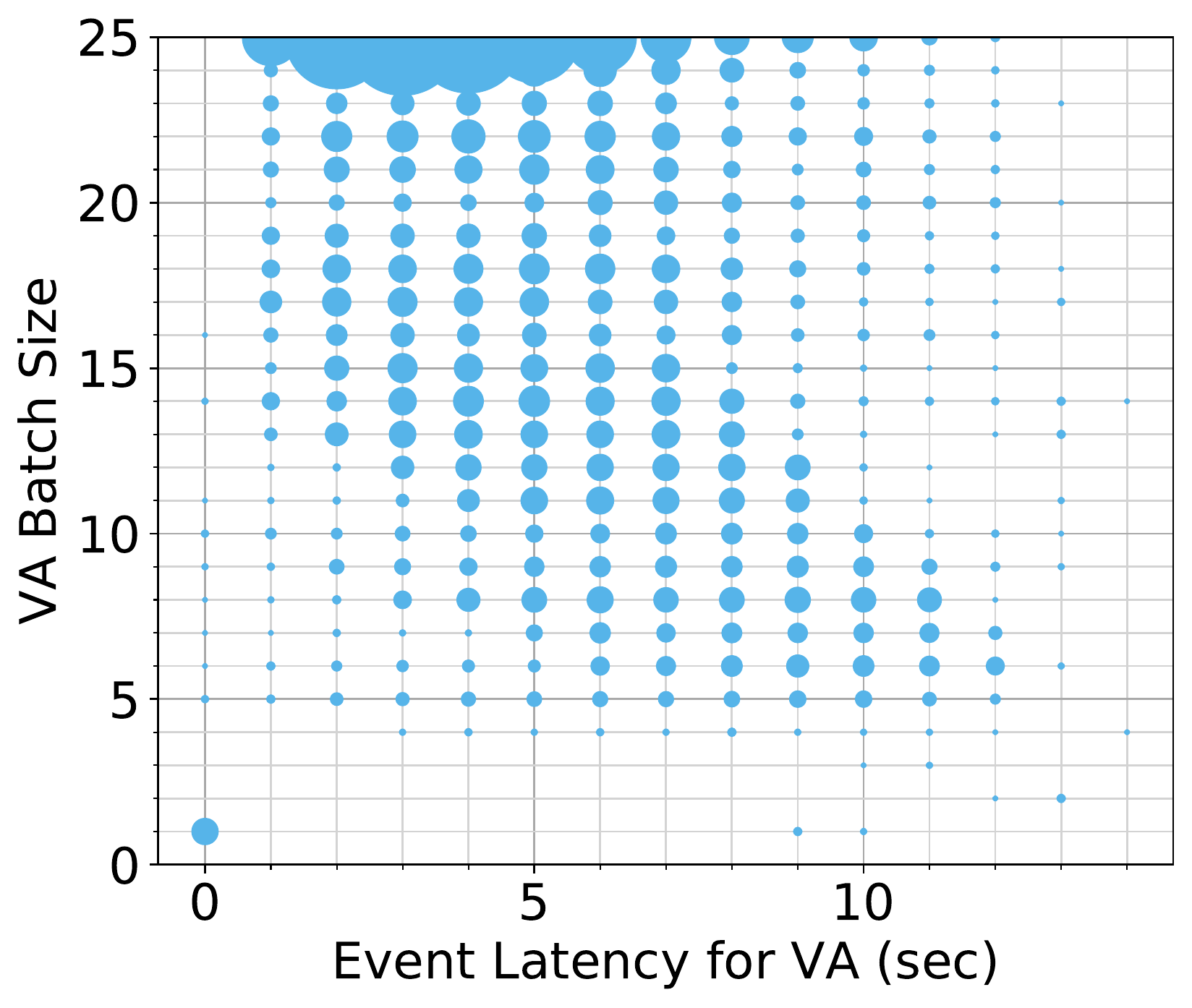}
	}
	\subfloat[CR Batch size vs. Event latency]{\label{fig:cr_lat_4ms}
		\includegraphics[width=0.47\columnwidth]{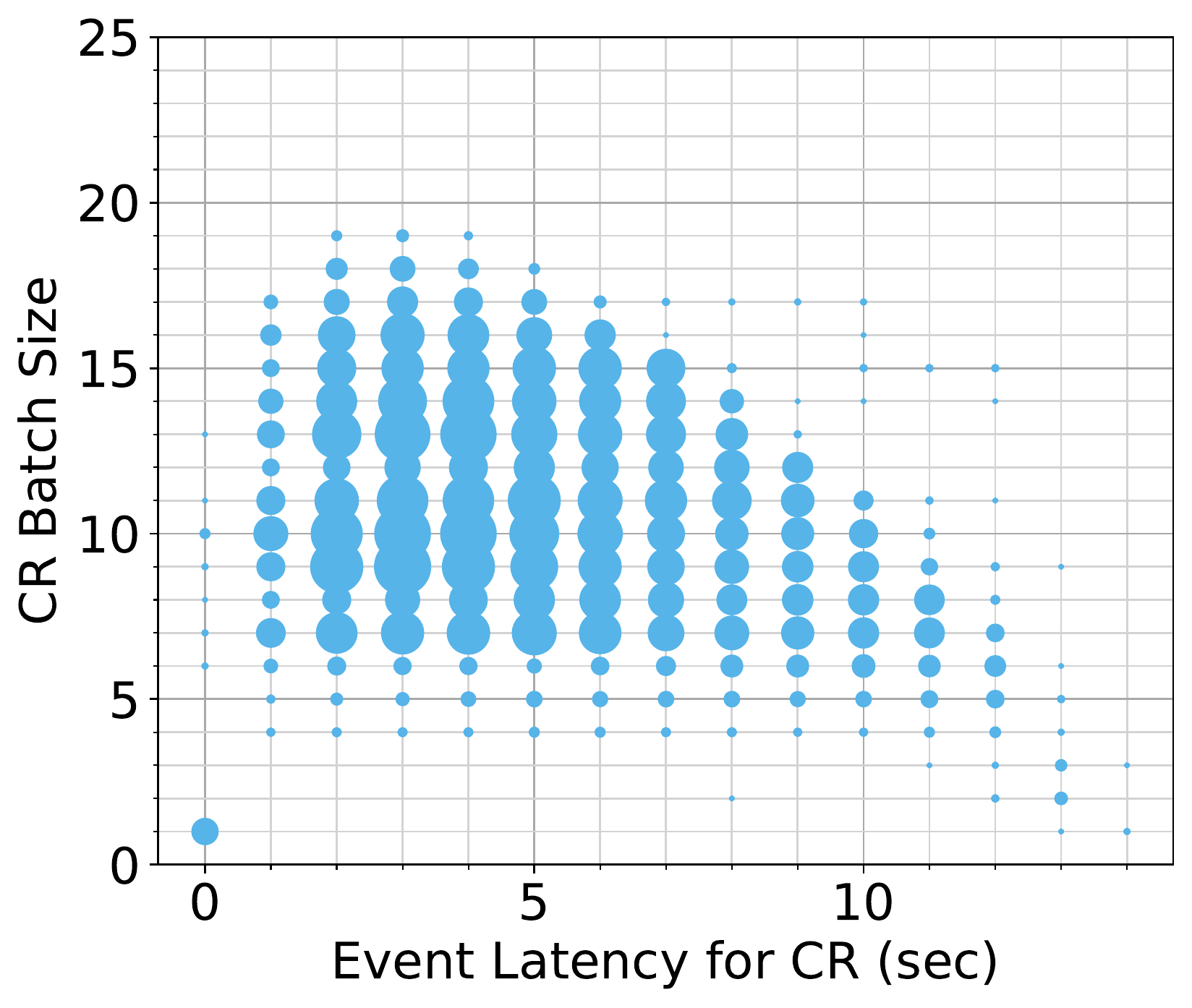}
	}
	\caption{Performance of \anv batching, TL-BFS, \emph{es=4~m/s}}
\end{figure}

\para{Analysis of \anv's Batching}
We further analyze the behavior of \anv's dynamic batching by examining the two key tasks that dominate the execution latency, VA and CR. The dynamic batch sizing operates independently for these two tasks types.

Figs.~\ref{fig:4ms_dynamic_bs:va} and~\ref{fig:4ms_dynamic_bs:cr} show the active cameras count (left Y axis) and the batch size averaged every $1~sec$ (right Y axis), for all $10$ VA and CR tasks % mean VA batch sizes, averaged over every second of the application's execution time. The X axis represents the
for the application execution time-time (X Axis), while Figs.~\ref{fig:va_lat_4ms} and~\ref{fig:cr_lat_4ms} show a bubble (scatter) plot of the task latency per event against the batch size the event was part of, for all VA and CR instances.

For the \emph{VA task}, we see from Fig.~\ref{fig:4ms_dynamic_bs:va} that the batch size increases as the active camera count grows, helping it support a higher input event throughput. VA uses almost every batch size between 1 and 25 (Fig.~\ref{fig:va_lat_4ms}), and the latency varies within a single batch size -- events in a batch that arrive later will have a lower latency, and vice versa. For the larger batch sizes, the task latency often ranges from 2--6~secs, indicating that the VA module can benefit from a more relaxed $b^{max}$.

The \emph{CR task} shows a similar trend between the camera count and batch size in Fig.~\ref{fig:4ms_dynamic_bs:cr}. CR has a lower mean batch size than VA, which is expected since the CR module is a compute intensive DNN which has a larger execution time than the VA module.
Fig.~\ref{fig:4ms_dynamic_bs:cr} which is a timeline plot for mean number of CR shows a similar trend. Also, despite $b^{max}=25$, its peak batch size never exceeds 19. 

On further analysis, the maximum budget allocated to a CR task is $\beta=3.65~secs$. At the peak active camera count of 111, this task receives events from $13$ cameras. For forming a batch of size $b=25$, we have its queuing time as $1.92~secs$, assuming a uniform input rate of $13~events/sec$, and execution time of $\xi(25)=1.74~secs$, which together at $3.66~secs$ exceed the budget. Instead, the dynamic batch size selected of $b=19$ results in a processing time of $2.91~secs$, which is within the budget. This indicates that \anv's dynamic batch sizing is sensitive to the needs of individual tasks well.
In fact, even with a peak entity speed of $es=6~m/sec$, it is able to avoid any event delays (Fig.~\ref{fig:bar_6ms}, DB-25).

\begin{figure}[t]
	\centering
	\subfloat[NOB Batching ($b^{max}$=$25$)]{\label{fig:4ms_30Mbps_down_base}%NOB | 300s-1G:300s-30M 
		\includegraphics[width=0.47\columnwidth]{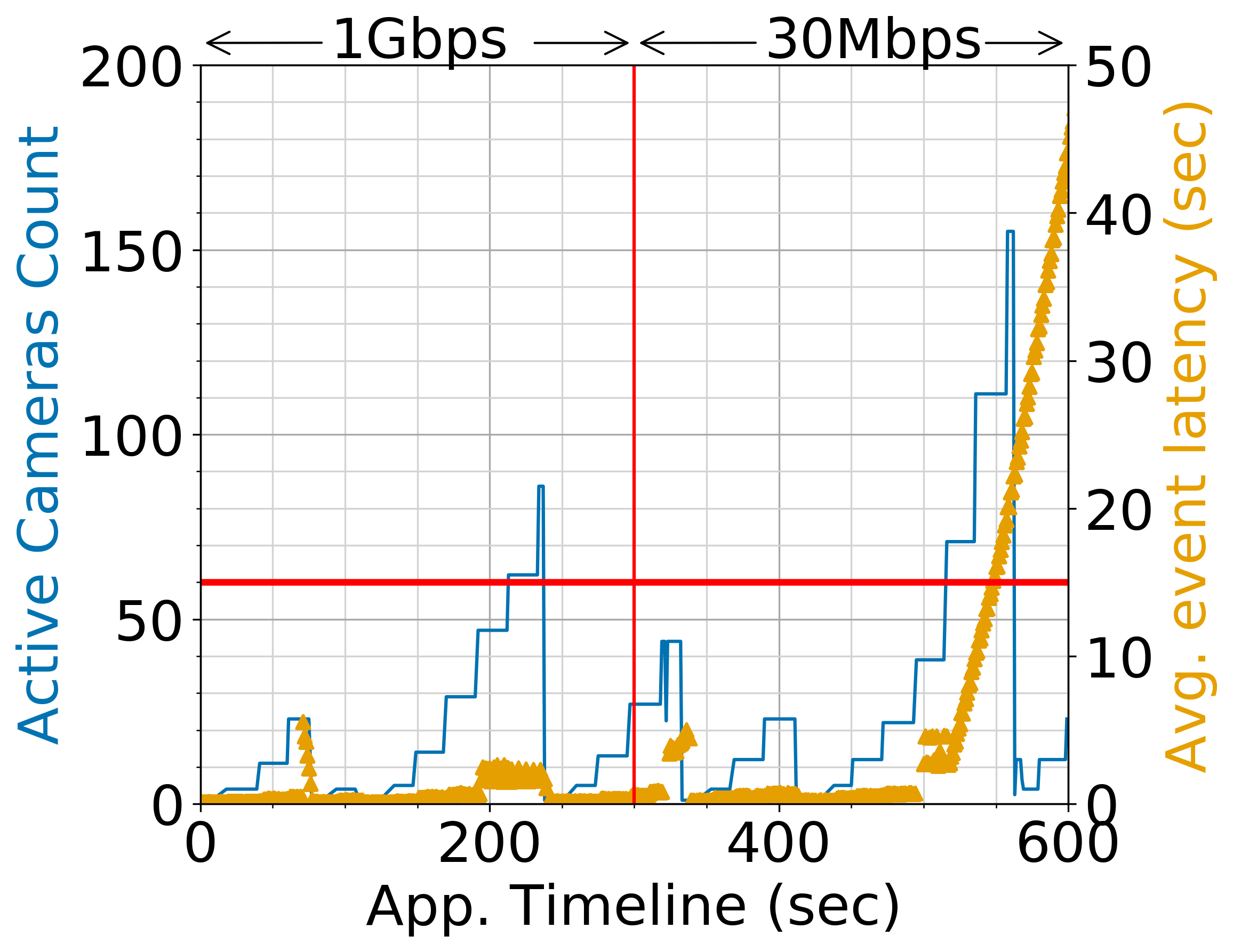}
	}~~
	\subfloat[\anv Batching ($b^{max}$=$25$)]{\label{fig:4ms_30Mbps_down}%300s-1G:300s-30M
		\includegraphics[width=0.47\columnwidth]{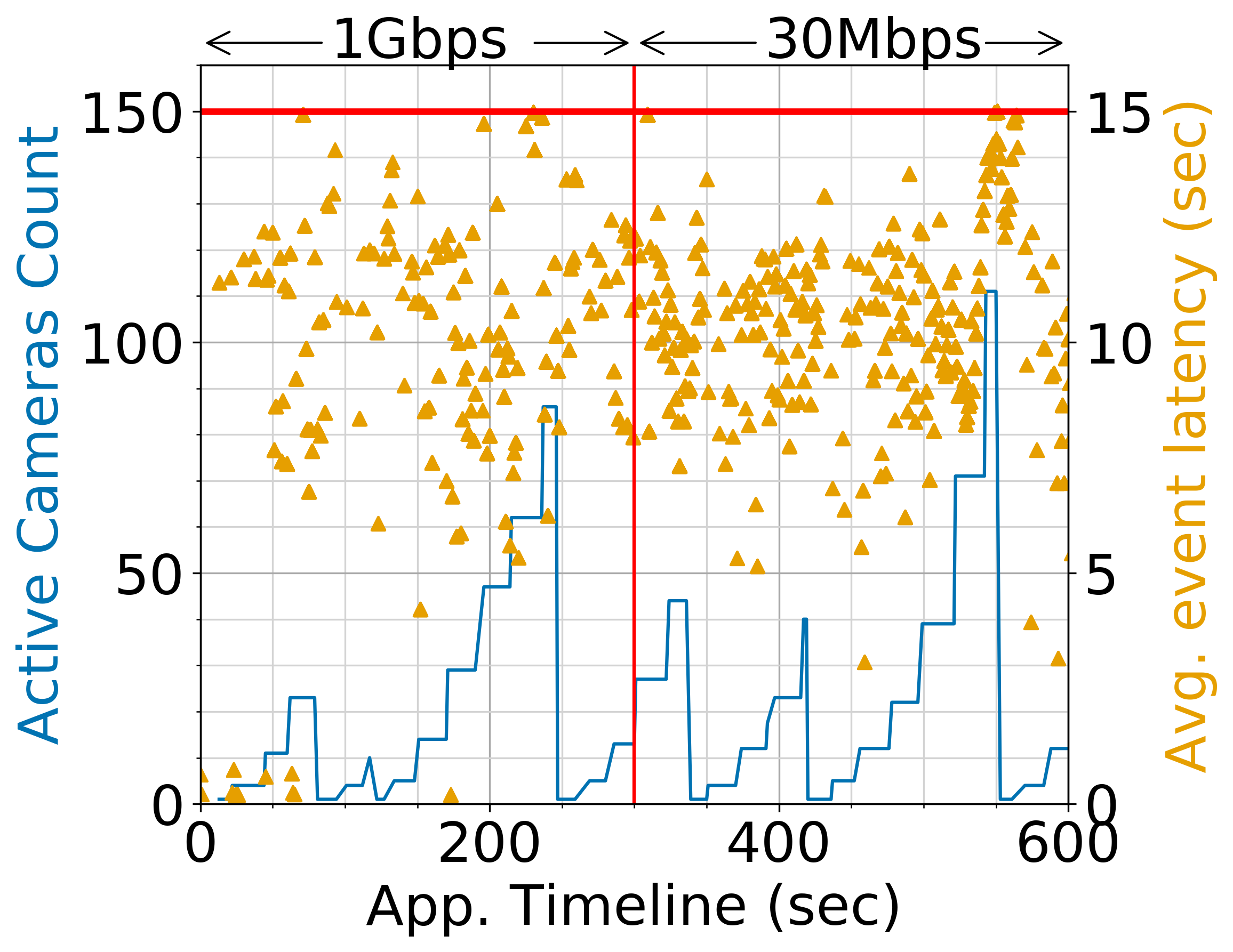}
	}
	\caption{Adapting to network variation. The system bandwidth drops from $1~Gbps$ to $30~Mbps$ after the $300^{th}~sec$.}
	\ysnote{Add a header line above LHS and RHS of plot indicating ``1Gbps'' and ``30Mbps'', for both plots}
\end{figure}
\para{Adapting to network variation}
The complexity of \anv's batching logic is partly attributed to its ability to respond to handling network and computation variability. The former is more common in WAN and MAN.  We evaluate its ability to adapt to even sharp changes in the network performance. Using the dynamic setup for \anv and NOB from Figs.~\ref{fig:4ms_base_dynamic} and~\ref{fig:4ms_dynamic}, we drop the bandwidth between compute nodes from $1~Gbps$ to $30~Mbps$ midway through the application execution at $300~secs$ into the timeline. These are shown in Figs.~\ref{fig:4ms_30Mbps_down_base} and~\ref{fig:4ms_30Mbps_down} for \anv and NOB. 

The first $300~secs$ is identical to the earlier plots, and neither configuration has event delays. But once the bandwidth drops, \anv manages to keep the system stable with no event delay as it reacts to event latencies increasing. As the network degrades, the budget available to tasks reduce, causing smaller batch sizes to be formed. The median CR batch size rapidly drops from $b=8$ to $5$, and the batches with $1$ and $2$ events rise from only $\approx18\%$ before $300~secs$ to $\approx30\%$ after the network slowdown.
However, NOB becomes unstable and its event latency grows beyond $\gamma$ after $500~secs$. This is due to its lookup table being created for a certain system performance and that not holding at runtime.

\subsubsection{Analysis of Tracking Logic}

We next consider the benefits of the tracking logic knob of the tuning triangle in helping control the active camera count and scalability of \anv. We compare the prior \emph{TL-BFS} logic with two others, \emph{TL-Base} that is simpler, and \emph{TL-WBFS} that is more advanced. We use static batching, keep drops disabled and $es=4~m/sec$.

\begin{figure}[t]
	\centering
	\subfloat[TL-WBFS ($b=1$)]{\label{fig:4ms_tl-wbfs}
		\includegraphics[width=0.52\columnwidth]{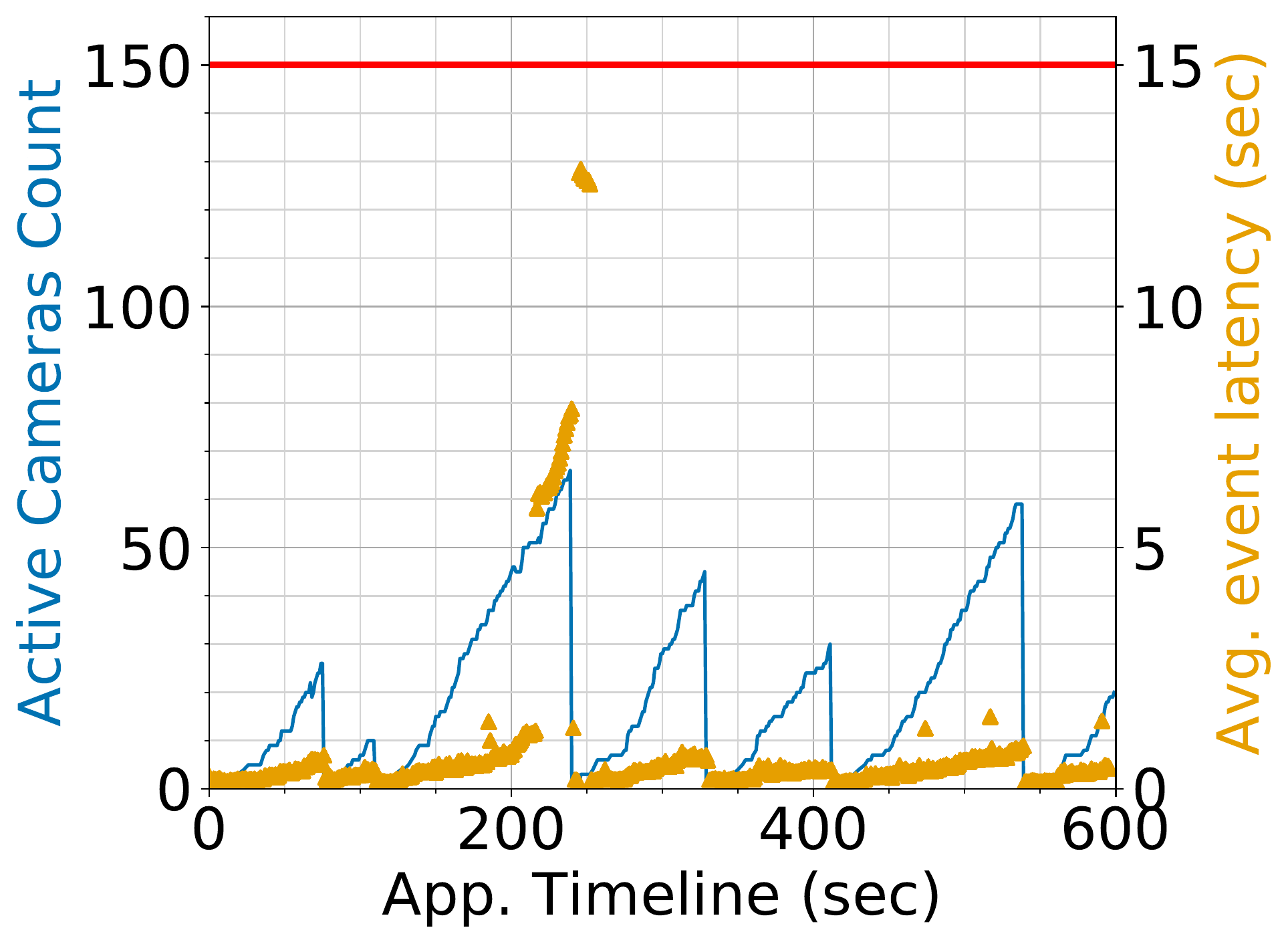}
	} 
	\subfloat[TL-Base ($b=20$)]{\label{fig:4ms_tl-base}
		\includegraphics[width=0.46\columnwidth]{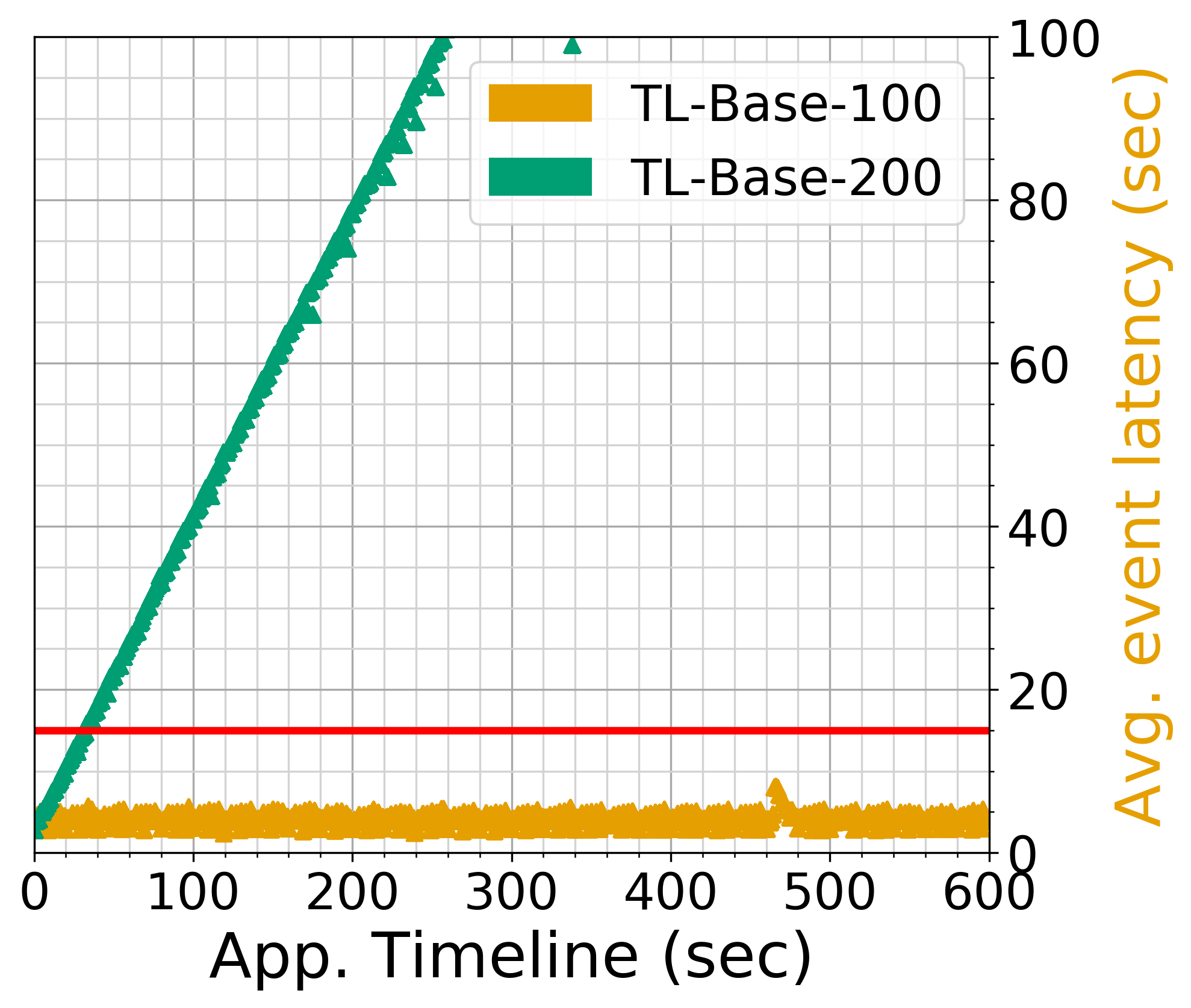}
	}
	\caption{Effect of \emph{tracking logic} on performance, \emph{es=4~m/sec}}
\end{figure}

\emph{TL-Base} is a baseline logic that keeps all cameras active, similar to contemporary systems. Since the resources are inadequate to support all $1000$ cameras being active, we do two runs, with $100$ and $200$ cameras placed on a proportionally smaller road network, and all active. We use a static batch size of $b=20$, which offers the best configuration. Fig.~\ref{fig:4ms_tl-base} plots the application time-line and the event latency averaged over $1~sec$ for the 100 and 200 camera setup. While the event latency is stable without any delays for 100 cameras, it is unstable and grows rapidly with 200 active cameras, indicating a lack of resources. The total frames processed is $\approx 60,000$ in the former, and $\approx 120,000$ in the latter with over $55\%$ delayed (Fig.~\ref{fig:bar_4ms}, Base SB-20 100c and 200c).
Obviously, this does not scale to $1000$ cameras, unlike our runs with better TL algorithms.

The more advanced TL strategy \emph{TL-WBFS} supports 1000 total cameras on the same set of resources by being smarter about which ones are activated. It supports a stable latency with $b=1$ streaming (Fig.~\ref{fig:4ms_tl-wbfs}) with no events delayed (Fig.~\ref{fig:bar_4ms}, WBFS SB-1), in contrast to TL-BFS which was unstable for $b=1$ (Fig.~\ref{fig:streaming}). The active camera count grows in more granular steps using WBFS since it is aware of the road lengths and leads to a more measured growth of active cameras.  % vertices in TL-WBFS are not equidistant to their childern. Also, when TL-WBFS is used, 
Further, its peak active camera count is $67$, relative to $111$ when using TL-BFS. So WBFS can help scale to a larger set of total cameras or for a longer period of the entity being in a blindspot. 

While a better TL helps, it is not a substitute for dynamic batching since we can have scenarios where a static batch is not adequate. E.g., for a faster $es=6~m/sec$ , TL-BFS with a static batch size of $b=20$ causes $603$ events to be delayed, compared to no delays when using dynamic batching (Fig.~\ref{fig:bar_6ms}, BFS SB-20 vs. DB-25).

\subsubsection{Analysis of Dropping Strategy}
Even TL and dynamic batching may not suffice when the spotlight grows large. This can cause the resources to be overwhelmed, latencies to grow unabated, and cascade to all future events. \anv's smart dropping strategy is beneficial here to drop events early in the pipeline to avoid resource wastage, and reduce overall event delays.

\begin{figure}[t]
	\centering
	\subfloat[Drops Disabled]{\label{fig:7ms_dynamic batching}
		\includegraphics[width=0.48\columnwidth]{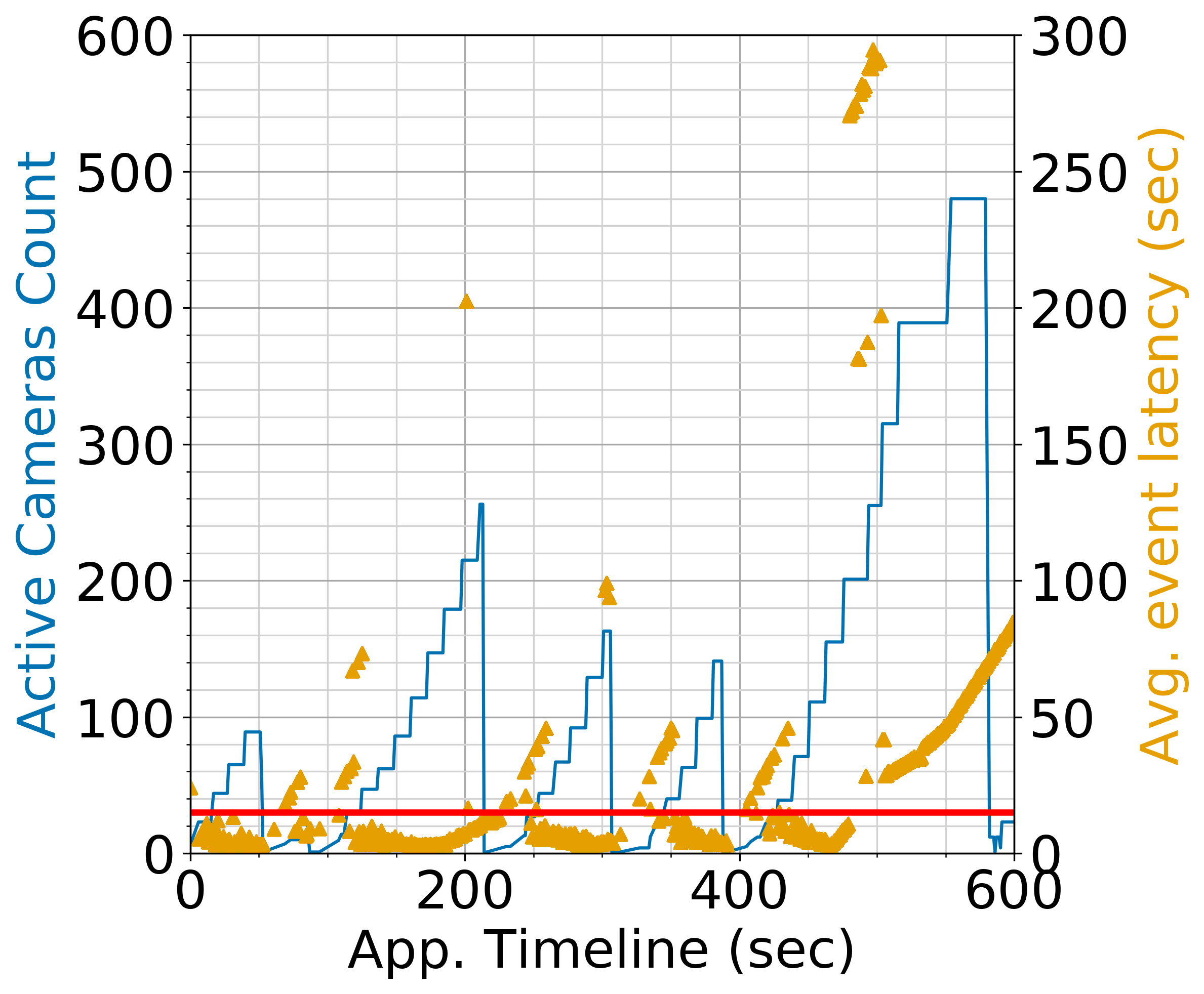}
	}
	\subfloat[Drops Enabled]{\label{fig:7ms_dynamic_drops}%\label{fig:7ms_dynamic}
		\includegraphics[width=0.52\columnwidth]{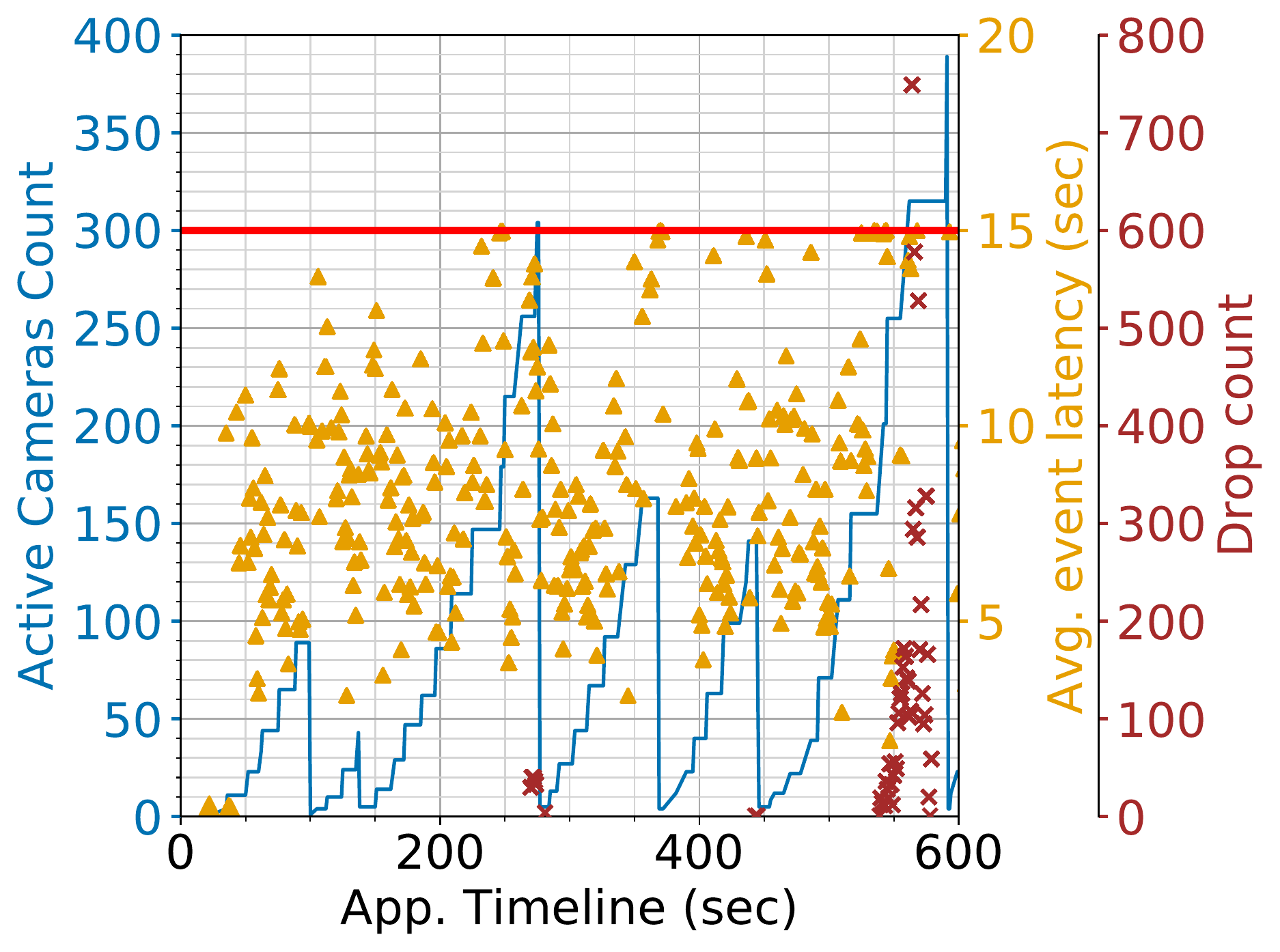}
	}\\
	\caption{Perf. with \emph{drops dis/enabled}, \emph{TL-BFS} and \emph{es=7~m/sec}} 
	\ysnote{increase height of (a) retaining current width}
\end{figure}

We run the App 1 experiments from \S~\ref{sec:exp:batching}, using our dynamic batching and TL-BFS, but with a faster peak walk speed of $es=7~m/s$. This causes the spotlight to grow faster when the entity is in a blindspot. Under such conditions, when \emph{drops are disabled}, we see from the timeline plot in Fig.~\ref{fig:7ms_dynamic batching} that the application is unstable, with latency $\gg \gamma$ as the active cameras grow from 100--500. This causes each CR instance to receive a peak of $\approx 49~events/sec$, while its capacity is only $19~events/sec$. This results in $85\%$ of events to be delayed (Fig.~\ref{fig:bar_7ms}, DB-25).

When \emph{drops are enabled}, the application's latency is stable and within $\gamma=15~secs$ even when the active camera count grows as high as 389 (Fig.~\ref{fig:7ms_dynamic_drops}). The drops start when the camera count exceeds 200, which matches $\approx 20~events/sec$ for each CR task. While $17\%$ of all events are dropped, the rest of the events are processed without any delays (Fig.~\ref{fig:bar_7ms}, DB-25 Drops). Dropping frames containing the entity can delay locating the entity and cause the active set to grow. However, none of the $21$ frames carrying the detected entity is dropped. This is merely incidental, but enabling the \emph{do not drop} flag will actually ensure this. The batch sizes for VA and CR are smaller here (not shown) than for $es=4~m/sec$ with dynamic batching but no drops (Figs.~\ref{fig:va_lat_4ms} and~\ref{fig:cr_lat_4ms}). When the input event rate is high, the system first reduces the queuing time by forming smaller batches and then resorts to drops. 

\subsection{Analysis of App 2}
\begin{figure}[t]
	\centering
	\subfloat[Latency, \emph{4m/sec}]{\label{fig:app2:lat_4ms}
		\includegraphics[width=0.25\columnwidth]{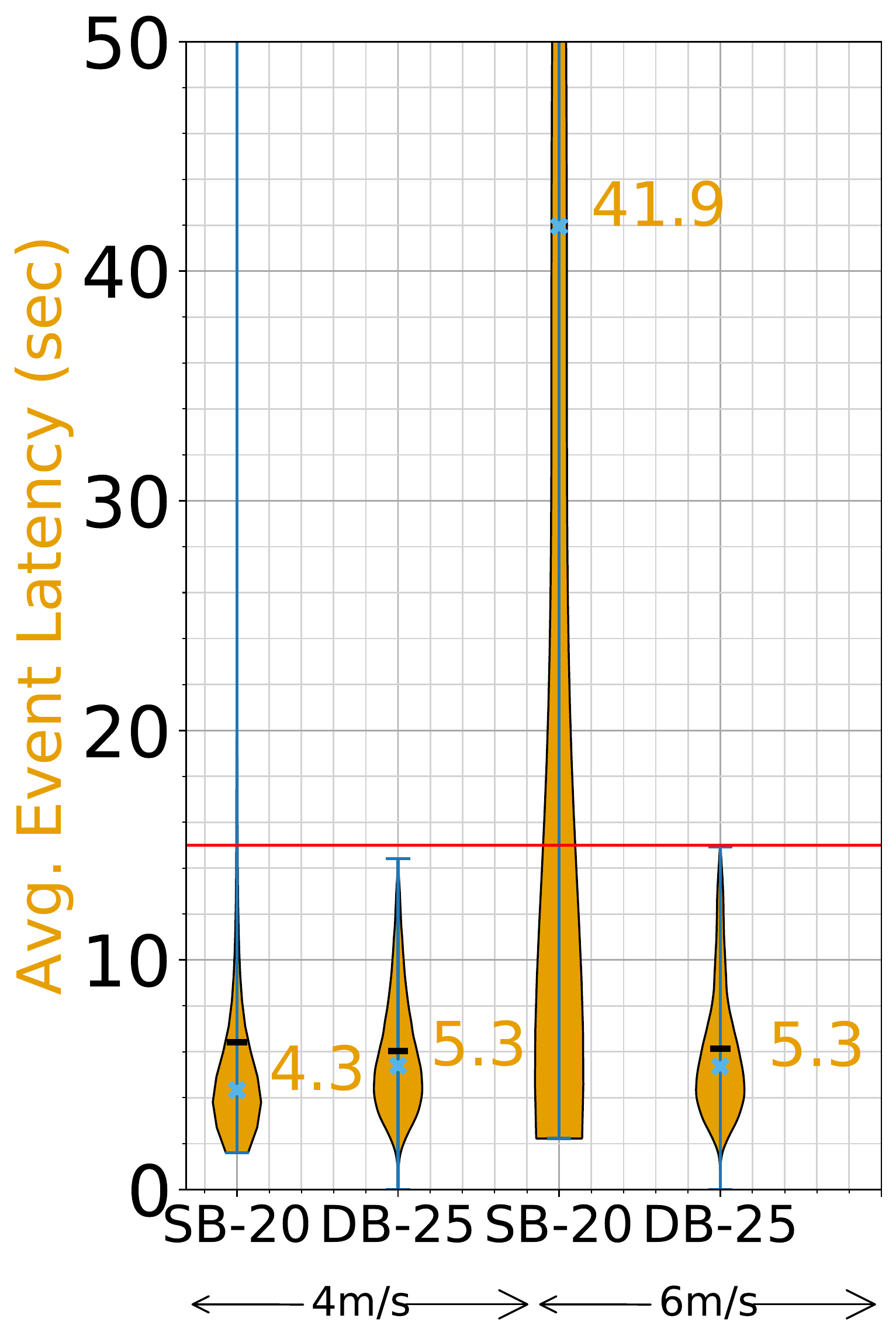}
	}
	\subfloat[Delays, \emph{4m/sec}]{\label{fig:app2:bar_4ms}
		\includegraphics[width=0.25\columnwidth]{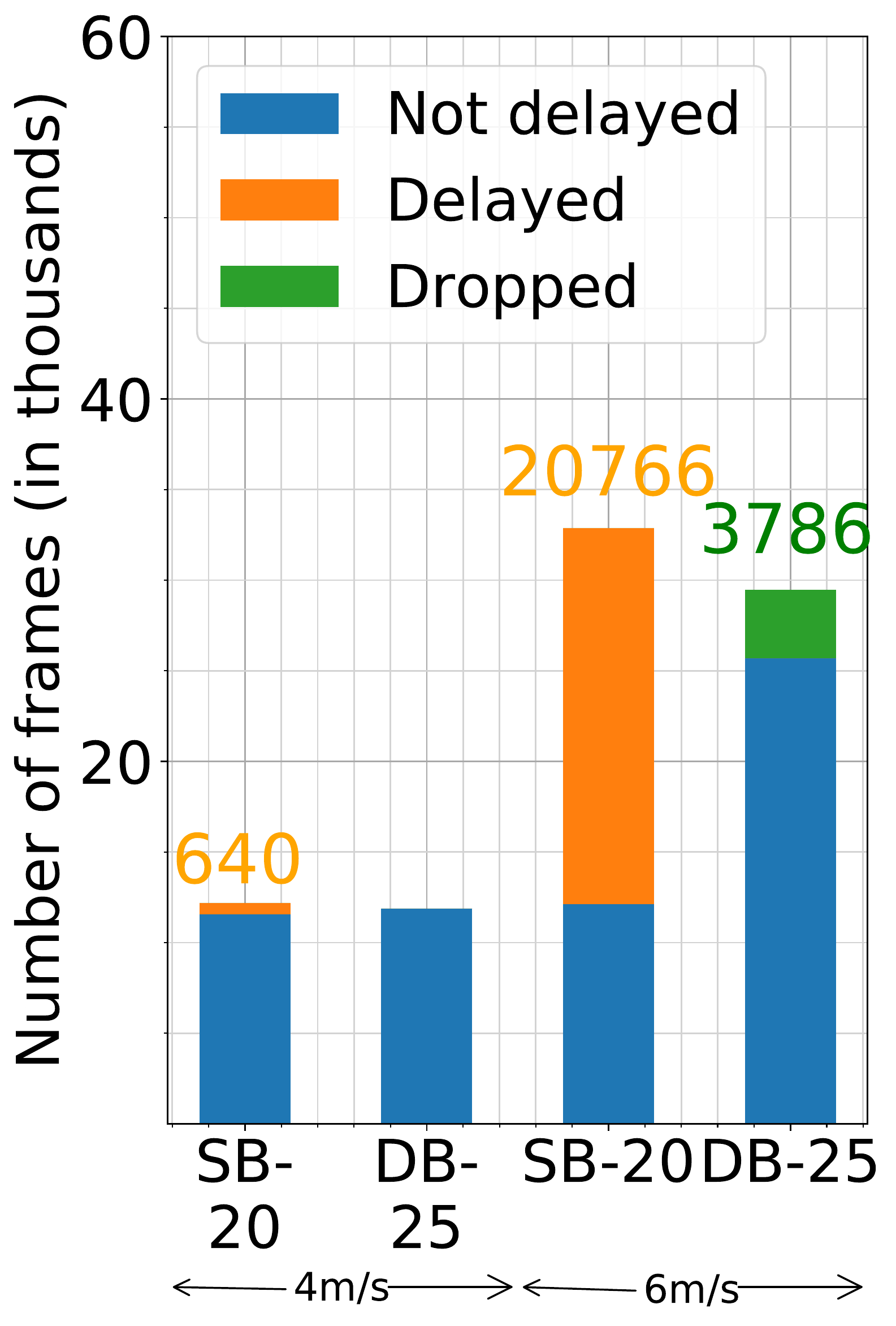}
	}~
	\subfloat[Active camera count for App 1 vs. App2 over time]{\label{fig:app1_vs_app2}
		\includegraphics[width=0.45\columnwidth]{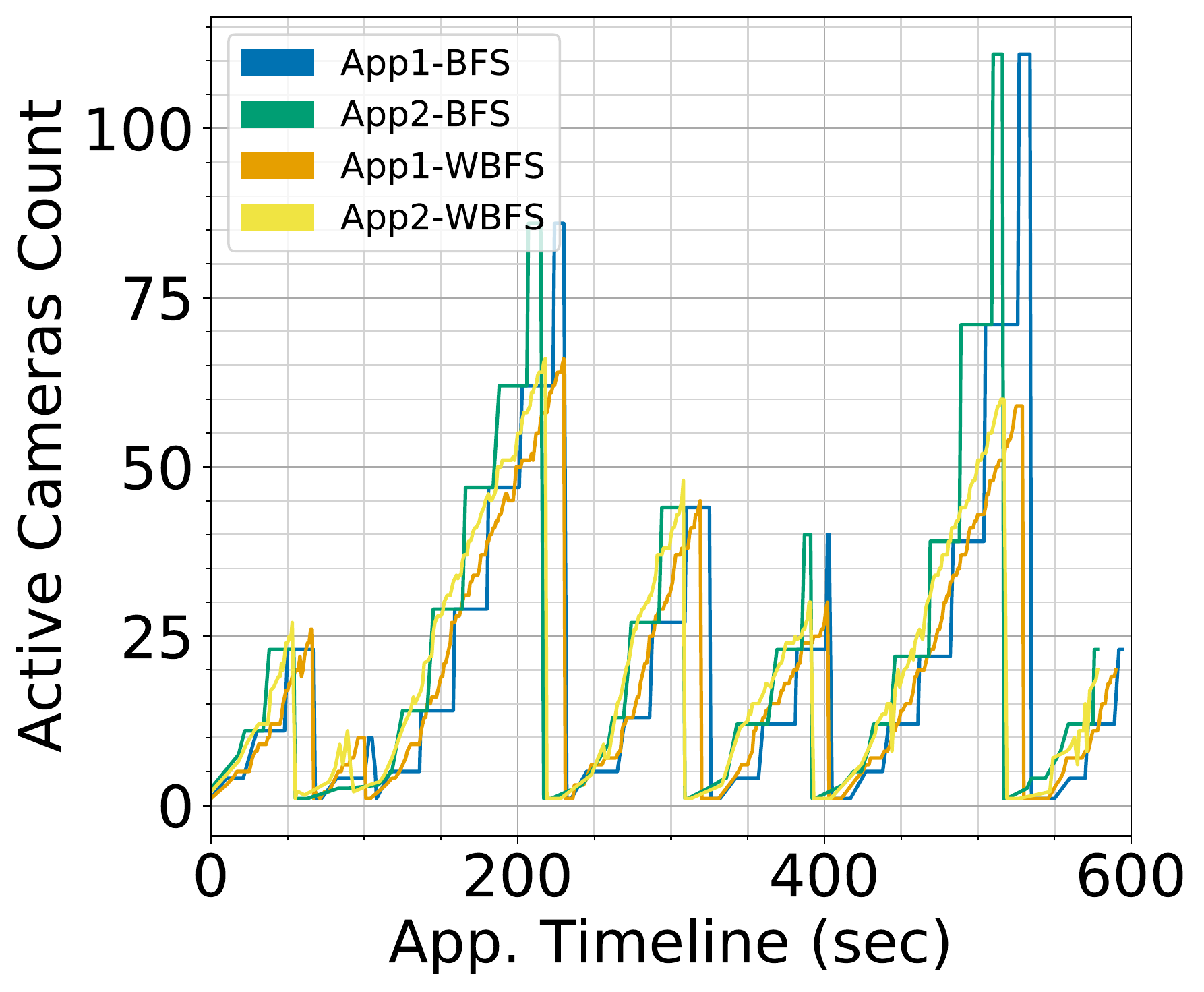}
	}
	\caption{Latency distribution, event delays and camera count for different configurations of \emph{App 2}, using \emph{TL-BFS}}\label{fig:app2}
	\ysnote{retaining the current width, increase height of (a) and (b) to match (c). Y axis of (b) to stop at 60. Increase font sizes for (a-c). You can drop ``es='' and just say ``4m/s'' to save space.}
\end{figure}

In this section, we perform a subset of the above experiments for App 2 to reconfirm the tuning triangle trends. They key difference between these two applications is CR, with the logic for App 2 using a more accurate and compute-intensive DNN that takes $\approx63\%$ longer to process each frame than for App 1. We use the same road network, entity query, 1000-camera setup, $\gamma=15~secs$, BFS tracking logic, drops disabled and peak entity speed of $es=4~m/s$ by default, unless mentioned otherwise. Figs.~\ref{fig:app2} show the \emph{latency distribution} and the \emph{number of delayed events} for App 2, for various runs, while Fig.~\ref{fig:app1_vs_app2} shows the \emph{active camera count} for App 1 and App 2 using BFS and WBFS TL along the application timeline. 

Using a static batch size of $b=20$, we observe a median latency of $4.33~secs$ but with $\approx 5\%$ latency violation (Figs.~\ref{fig:app2:lat_4ms} and~\ref{fig:app2:bar_4ms}, BFS SB-20). But with dynamic batching enabled with $b^{max}=25$, we see a median latency of $5.39~secs$ but crucially, no latency delays for events (Figs.~\ref{fig:app2:lat_4ms} and~\ref{fig:app2:bar_4ms}, BFS DB-25). This confirms the need for and benefits of \emph{dynamic batching} in App 2 as well.

The tracking logic in App 2 plays an important role in managing the growth in the active camera set size, similar to App 1. In Fig.~\ref{fig:app1_vs_app2}, using a static batch size of $b=20$, we see that TL-WBFS, which uses the knowledge of road lengths for its spotlight expansion, has a more modest increase in camera count, e.g., at $\approx 500~secs$, compared to TL-BFS. This help it scale to a denser camera deployment or a longer duration of the entity being in a blindspot.  Both App 1 and App 2 have similar camera count patterns, partly modulated by the different CRs used.

Finally, for $es=6m/s$ we see that dynamic batching is inadequate for TL-BFS as it reports a median latency of $41.95~secs$ and $63\%$ of events delayed (Figs.~\ref{fig:app2:lat_4ms} and~\ref{fig:app2:bar_4ms}, BFS DB-25). But by enabling drops, this reduces the drops to $\approx12\%$ with other events being processed within time, with a median latency of $5.36~secs$ (BFS DB-25 Drops). Here again, App 2 exhibits the value of \emph{pro-active and intelligent drops} by \anv rather than allow delayed events to flow through and waste resources.

\section{Related Work}
\label{sec:related}
\para{Video Surveillance Systems} 
Intelligent video surveillance systems allow automated analytics over camera feeds~\cite{al2014video}. They enable wide-area surveillance from multiple static or PTZ cameras, with distributed or master-slave controls~\cite{survey-tmcca-2015}. 
\emph{ADVISOR}~\cite{siebel2004advisor} supports tracking, crowd counting, and behavioral analysis over camera feeds from train stations to support human operators. However, these are pre-defined applications, run centrally on a private data center and process all camera feeds all the time. \emph{IBM's Smart Surveillance System (S3)}~\cite{shu2005ibm} is a proprietary platform for video data management, analysis and real-time alerts. While it offers limited composability using different modules, it too executes the applications centrally and does not consider performance optimizations. Early works examine edge computing for basic pre-processing~\cite{kornecki2008middleware}. But the edge logic is statically defined, with the rest of the analytics done centrally and over dedicated networks. 

The \emph{Ella middleware}~\cite{dieber2013ella} supports the definition of distributed analytics on the edge and over a WAN. They use a publish-subscribe model with hierarchical brokers to route video and event streams between analytics deployed on edge devices. They illustrate a multi-camera single person tracking application, similar to us. However, their platform design resembles a general-purpose event-driven middleware, without any specific model support or runtime optimizations for video analytics, unlike us. Others exclusively focus on offline analysis over video feeds from a many-camera network along with other data sources for spatio-temporal association studies~\cite{smartmon-tbd-2018}.

\emph{Vigil}~\cite{Zhang:2015:DIW:2789168.2790123} is designed for video surveillance on wireless networks with limited bandwidth. They assume an Edge computing node (ECN) is co-located with the cameras and is used to reduce redundant data from being sent to the Cloud. The authors assign a utility score to each frame to ascertain its importance, similar to our \emph{do not drop} flag. Our model and platform offer more active controls over the logic running on the ECN, and the runtime tuning. 

\emph{EdgeEye}~\cite{edgeeye} efficiently deploys DNN models on the edge using a JavaScript API for users to specify their parameters. While it caters to a wider class of analytics applications, it lacks composability and domain-specific patterns for tracking applications. It offers performance optimizing for the DNN model, but does not consider distributed systems issues, such as batching, dropping and network variability. 
\emph{Video Storm}~\cite{zhang2017live} is a video analytics system with the goals of approximation and delay tolerance. It schedules video analytics query workloads on a cluster of machines, where each query has a deadline and priority. \emph{Video Edge}~\cite{videoedge-sec-2018} extends this to support scheduling on a hierarchy of edge, fog and cloud resources. Both these provide tuning knobs which are conceptually similar to our ours. But the key distinction is that these degrees of freedom requires the specification of objective functions to define the impact of the knobs on metrics. This makes it challenging to use out of the box. Our domain-sensitive Tuning Triangle more intuitively captures the impact of the three well-defined knobs on the three metrics that have the most impact on tracking applications. 

While we propose and demonstrate several graph-based tracking logic, probability -based and spatio-temporal models have been explored in literature. Jain, et al.~\cite{jain2018rexcam} perform an empirical evaluation of spatio-temporal constraints and correlations, and leverage these to improve the precision and reduce computational cost of cross-camera video analytics. Others~\cite{shiva2017distributed} perform offline analysis to learn the spatio-temporal constraints between cameras. During the online tracking phase, these are used to prune the number of potential paths. Kalman Filter has also been used to estimate the motion probability of the object being tracked~\cite{kang2005persistent} . 

\para{Big Data platforms and DSL}
Generic stream processing platforms like \emph{Apache Storm, Flink} and \emph{Spark Streaming}~\cite{spark,toshniwal2014storm,carbone2015apache} offer flexible dataflow composition. But defining a domain-specific dataflow pattern for tracking applications, like we do, offers users a frame of reference for designing distributed video analytics applications, with modular user-defined tasks. 

\emph{Google's \tf}~\cite{tensorflow} is a DSL for defining DNNs and CV algorithms, and to deploy trained models for inference. However, \tf is not meant for composing arbitrary modules together. The tasks take a Tensor as an input and give a Tensor as the output, and there are no native patterns such as Map and Reduce to ease composability. Yahoo's \emph{\tf on Spark}~\cite{tfonspark} gives more flexibility by allowing Spark's Executors to feed RDD data into \tf. Thus, users can couple Spark's operations with \tf's neural networks. But \anv is at a level of abstraction higher, allowing for rapid development of tracking applications with fewer lines of code or sometimes just a configuration change. Also, Spark is not designed for distributed computing on  WANs or edge/fog devices, which we address in the \anv runtime. 

\para{Streaming Performance Management}
There are several performance optimization approaches adopted by stream processing systems, which we extend. \emph{Apache Flink}~\cite{carbone2015apache} and \emph{Storm}~\cite{toshniwal2014storm} support \emph{back-pressure}, where a slow task sends signals to its upstream task to reduce its input rate. This may eventually lead to data drops, but the data being dropped are the new ones generated upstream rather than the stale ones that are already in-flight, which sacrifices freshness in favor of fairness. Our drops prioritize recent events over stale events, and also adjust the budget more precisely.

\emph{Google's Millwheel}~\cite{akidau2013millwheel} uses the concept of \emph{low watermarks} to determine the progress of the system, defined as the timestamp of the oldest unprocessed event in the system. It guarantees that no event older than the watermark may enter the system. Watermarks can thus be used to trigger computations such as window operations safely. While our batching and drop strategies are similar, watermarks cannot budget the time left for a message in the pipeline and has no notion of user-defined latency.

\emph{Aurora}~\cite{abadi2003aurora} use the concept of \emph{load shedding}, which is similar to our data drops. They define QoS as a multidimensional function, including attributes such as response time, similar to our maximum latency. Given this function, the objective is to maximize the QoS. \emph{Borealis}~\cite{abadi2005design} extends this to a distributed setup. \anv uses multiple drop points even within a task, which offers fine-grained control and robustness. Features like ``do not drop'' and resilience to clock skews are other domain and system-specific optimizations. 

\section{Conclusions}
\label{sec:conclusions}
In this paper, we have proposed an intuitive domain-specific dataflow model for composing distributed object tracking applications over a many-camera network. Besides offering an expressive and concise pattern, we surface the Tracking Logic module as a powerful abstraction that can perform intelligent tracking and manage the active cameras. This enhances the scalability of the application and makes efficient use of resources. 
Further, we offer tunable runtime strategies for dropping and batching that help users easily balance between the goals of performance, accuracy and scalability. Our design is sensitive to the unique needs of a many-camera tracking domain and for distributed edge, fog and cloud resources on wide-area networks. Our experiments validate these for a real-tracking application on feeds from up to 1000 cameras.

As future work, we plan to explore intelligent scheduling of the module instances on edge, fog and cloud resources; allow modules to be dynamically replaced for better accuracy or performance; handle mobile camera platforms such as drones. In a real setting, multiple objects of interest would be tracked across the camera network. This should lead to interesting scheduling problems as well as an opportunity to share compute across multiple queries.

\section{Acknowledgments} We thank Prof. A. Chakraborthy, Visual Computing Lab, IISc, for discussions on the tracking problem and video analytics modules. We also thank fellow students, Swapnil Gandhi and Anubhav Guleria, for their valuable insights. This work was supported by grants from Huawei as part of the Innovation Lab for Intelligent Data Analytics and Cloud, and resources provided by Microsoft and NVIDIA.

	\bibliographystyle{plain}
	\bibliography{arxiv}

\end{document}